\documentclass[manuscript,screen]{acmart}
\setcopyright{none}
\renewcommand\footnotetextcopyrightpermission[1]{} 
\AtBeginDocument{%
  }

\setcopyright{acmlicensed}
\copyrightyear{2018}
\acmYear{2018}
\acmDOI{XXXXXXX.XXXXXXX}
\acmConference[Conference acronym 'XX]{Make sure to enter the correct
  conference title from your rights confirmation email}{June 03--05,
  2018}{Woodstock, NY}

\acmISBN{978-1-4503-XXXX-X/2018/06}

\usepackage{geometry}
\usepackage{graphicx}
\usepackage{float}
\usepackage{wrapfig}

\usepackage{multirow}
\usepackage{array}
\usepackage{longtable}
\usepackage{booktabs}
\usepackage{pdflscape}

\usepackage{amsmath}

\usepackage{amssymb}

\usepackage{natbib}
\usepackage{hyperref}
\usepackage{cleveref} 
\usepackage{tcolorbox} \tcbuselibrary{skins}

\hypersetup{
    colorlinks=false,
    bookmarksnumbered=true,
    bookmarksopen=false,
    pdftitle={Theory-Informed Messages for Remote Assessment},
    pdfauthor={Authors}
}

\begin{document}

\title[Theory-Informed Messages to Balance Cheating, Performance and UX]{Can Theory-Informed Message Framing Drive Honest and Motivated Performance with Better Assessment Experiences in a Remote Assessment?}

\author{Suvadeep Mukherjee}
\email{suvadeep.mukherjee@uni.lu}
\orcid{0000-0002-1213-1767}
\affiliation{%
  \department{Interdisciplinary Centre for Security, Reliability and Trust (SnT)}
  \institution{University of Luxembourg}
  \city{Kirchberg}
  \country{Luxembourg}
}
\additionalaffiliation{%
  \department{Department of Behavioural and Cognitive Sciences}
  \institution{University of Luxembourg}
  \city{Belval}
  \state{Esch-sur-Alzette}
  \country{Luxembourg}
}

\author{Björn Rohles}
\affiliation{%
  \institution{Digital Learning Hub}
  \city{Belval}
  \country{Luxembourg}
}
\email{info@rohles.net}

\author{Gabriele Lenzini}
\affiliation{%
  \department{Interdisciplinary Centre for Security, Reliability and Trust (SnT)}
  \institution{University of Luxembourg}
  \city{Kirchberg}
  \country{Luxembourg}
}
\email{gabriele.lenzini@uni.lu}

\author{Pedro Cardoso-Leite}
\affiliation{%
  \department{Department of Behavioural and Cognitive Sciences}
  \institution{University of Luxembourg}
  \city{Belval}
  \state{Esch-sur-Alzette}
  \country{Luxembourg}
}
\email{pedro.cardosoleite@uni.lu}

\renewcommand{\shortauthors}{Mukherjee et al.}

\begin{abstract}
Remote unproctored assessments increasingly use messaging interventions to reduce cheating, but existing approaches lack theoretical grounding, focus narrowly on cheating suppression while overlooking performance and experience, and treat cheating as binary rather than continuous. This study examines whether messages based on 15 psychological concepts from self-determination, cognitive dissonance, social norms, and self-efficacy theories can reduce cheating while preserving performance and experience. Through an expert workshop (N=5), we developed 45 theory-informed messages and tested them with online participants (N=1232) who completed an incentivized anagram task. Participants were classified as non-cheaters (0\% items cheated), partial-cheaters (1--99\% cheated), or full-cheaters (100\% cheated). Results show that concept-based messages reduced full-cheating occurrence by 42\% (33\%$\rightarrow$19\%), increased non-cheating by 19\% (53\%$\rightarrow$63\%), with no negative effects on performance or experience across integrity groups. Surprisingly, messages grounded in different theoretical concepts produced virtually identical effects. Analyses of self-rated psychological mechanisms revealed that messages influenced multiple mechanisms simultaneously rather than their intended targets, though these mechanisms predicted behavior, performance, and experience. These findings show that causal pathways are more complex than current theories predict. Practically, integrity interventions using supportive motivation rather than rule enforcement can reduce cheating without impairing performance or experience.

\end{abstract}








\maketitle

\section{Introduction}
Assessments are central to education and careers because they determine admission, certification, and advancement opportunities \cite{kam2018academic, ranger2020responding}. Their value depends on being fair, valid, and trustworthy so that assessment outcomes truly reflect students' competence and people can trust academic qualifications \cite{simons2019assessment}. When an assessment is compromised by dishonest practices, commonly known as `exam cheating', its fairness and validity are often undermined. Through cheating, dishonest students can gain unfair advantages, grades become inflated, and institutional credibility erodes \cite{alan2020cheating, alessio2017examining}. The studies observed that the challenge is even greater in remote contexts, where the absence of supervision and the ease of online access to exam responses make misconduct more tempting. Indeed, students themselves consistently report that it is easier to cheat in online tests than in traditional ones \cite{king2009online}, highlighting the urgent need for sustainable solutions.

To address the growing threat of exam cheating in remote settings, educational institutions have increasingly turned to remote proctoring tools, enabled by advances in technology and internet connectivity, that aim to replicate classroom supervision through digital surveillance \cite{li2021visual}. While these systems can effectively reduce exam cheating by increasing the perceived risk of being caught, they also create significant drawbacks including privacy intrusion, heightened stress and anxiety during exams, and potential performance impairment \cite{balash2021examining, chin2017influence, conijn2022fear}. Because of these concerns, many test institutions continue to rely on unproctored remote assessments, conducted without direct supervision and based on mutual trust between test-takers and institutions \cite{rios2017online}. Such assessments avoid surveillance-related problems but leave exams more vulnerable to cheating, creating perceptions of unfairness that can also erode trust in the assessment system \cite{daffin2018comparing, dendir2020cheating}. This dilemma has led to growing interest in \textbf{message-based interventions} as promising, low-cost alternatives without the invasive monitoring of proctoring systems. Institutions have experimented with various message-based interventions to reduce cheating, targeting either opportunities for dishonesty, fear of consequences, behavioral nudges, or moral reasoning. For example, some used honor code reminders to emphasize personal ethical responsibility \cite{bing2012experimental, mukherjee2023effects, tatum2022honor}, others issued warnings about the potential repercussions of misconduct \cite{corrigangibbs2015deterring, humbert2022role}, some employed fake surveillance cues to create a sense of being monitored \cite{mukherjee2023effects, pleasants2022cheating}, and others relied on nudges, such as positive reinforcement techniques, to encourage honest behavior \cite{henderson2024auditing, miller1975attribution}. These approaches are attractive because they are simple, scalable, and have showed substantial reductions in cheating, in some cases between 20\% and 70\%.

However, current message-based interventions suffer from four major shortcomings that limit both their effectiveness and wider adoption.

\textbf{1. Lack of theoretical grounding in message design:} Current message-based interventions are often developed ad hoc, relying on intuition rather than systematic theoretical frameworks. Much of the literature focuses on surface-level variations, such as when messages appear (before or during the exam; \cite{mukherjee2023effects}), how they are framed (as deterrent threats; \cite{humbert2022role}; or motivational nudges; \cite{lemaux2023honesty}), or whether combining message types is more effective than using a single one \cite{bing2012experimental, karim2014cheating, pleasants2022cheating}. These trial-and-error approaches have sometimes led to notable reductions in cheating, even by half, yet they provide limited insight into why interventions work. Because many messages (e.g., honor code) often blend multiple behavioral cues, such as moral appeals mixed with social norms or autonomy support, it remains unclear which specific psychological processes drive behavior change. Without isolating these processes, cumulative knowledge about what works, why it works, and under what conditions remains limited.

\textbf{2. Narrow focus on cheating suppression over holistic outcomes:} Most interventions are designed solely to reduce dishonest behavior, overlooking the need to balance assessment integrity with performance and experience. Effective interventions must preserve assessment validity by allowing test-takers to exhibit true competence, avoid adding cognitive burden through message content, and ensure a fair, positive testing experience \cite{zhao2024academic, mutimukwe2025privacy}. Moreover, research shows that academic performance and student wellbeing are interconnected and can influence each other reciprocally: students with greater wellbeing tend to achieve higher academic performance, while academic success can also enhance wellbeing \cite{vanderzanden2018domains}. However, very few studies examine this broader interplay among cheating prevention, performance factors, and user experience \cite{mukherjee2023effects}. This narrow focus limits both theoretical understanding and practical relevance for effective assessment design.

\textbf{3. Limited attention to motivational framing:} Many existing studies use messages that explicitly refer to cheating, honesty, or misconduct \cite{bing2012experimental, corrigangibbs2015deterring, humbert2022role, mukherjee2023effects, pleasants2022cheating}. Such framing can unintentionally make honest students feel accused or distrusted, potentially undermining their confidence in the testing environment \cite{mukherjee2023effects}. In contrast, messages that emphasize motivational factors, such as autonomy, competence, or descriptive norms, can effectively reduce cheating without mentioning dishonesty \cite{lois2021honest, pulfrey2019under}. This evidence suggests that cheating is shaped not only by opportunity or punishment but also by deeper motivational processes influencing performance and experience. When students are motivated by genuine interest, personal values, or a sense of ownership, they are less likely to cheat \cite{kanatmaymon2015role, murdock2006motivational}, perform better, and experience less stress and more positive emotions \cite{jian2019self, cao2023investigating}. Conversely, controlling or distrustful messages can undermine motivation, leading to disengagement and poorer performance \cite{vansteenkiste2018fostering}. Reframing message content around engagement, challenge, and personal growth has therefore potential to support both integrity and performance. For example, instead of saying ``It's up to you to stay honest while solving these challenges'', which implies distrust, a message like ``It's up to you how you tackle these challenges'' affirms autonomy and mastery. By aligning messages with broader motivational theories, interventions can move from reactive control to proactive engagement \cite{alqassab2024motivational, bennett2024beyond, kanatmaymon2015role}.

\textbf{4. Oversimplified conceptualization of cheating behavior:} Most prior studies treat cheating as a binary outcome, students either cheat or do not cheat \cite{corrigangibbs2015deterring, humbert2022role, mukherjee2023effects, pleasants2022cheating}. Research shows that test-takers typically exhibit distinct profiles\footnote{In this study, we call honest test-takers (0\% of words flagged as cheated) non-cheaters, occasional cheaters (1-99\% words flagged as cheated) partial-cheaters, and consistent cheaters (100\% words flagged as cheated) full-cheaters.}: occasional or partial-cheaters tend to balance self-image with material gain, consistent or full-cheaters act opportunistically when risks are low, and honest test-takers or non-cheaters rely on internalized norms \cite{pascualezama2020cheaters}. Critically, these groups respond differently to interventions. For example, moral reminders and self-concept nudges can reduce partial cheating but might have little effect on full-cheaters \cite{mazar2008dishonesty}, while norm-based strategies tend to reinforce honesty among those already honest \cite{cialdini2006managing}. These differential responses mean that universal intervention approaches risk being ineffective for some groups while potentially burdening others. Recognizing these distinct profiles allows examination of how theoretical concepts could motivate better performance differently across profiles, enabling practitioners to design tailored strategies for each group.

To address these shortcomings, this study develops a theoretical basis for designing messages that promote fair performance and a positive assessment experience while also reducing cheating. We draw on four well-established psychological frameworks, each targeting a different dimension of the assessment validity challenge. For example, self-determination theory emphasizes underlying mechanisms\footnote{The term ``mechanism'' refers to the psychological processes or mediators through which theoretical concepts influence motivation and behavior. For example, autonomy-need satisfaction \cite{ryan2000self, mcanally2024self}, normative perception \cite{cialdini2006managing}, or efficacy beliefs shaped by mastery and persuasion \cite{bandura1997self}.} such as psychological need satisfaction, particularly autonomy, competence, and relatedness, which foster volitional and confident performance \cite{ryan2000self}, something that surveillance can often undermine \cite{coghlan2021good}. Cognitive dissonance theory focuses on mechanisms related to self-concept consistency, helping test-takers resolve moral conflicts that might otherwise justify dishonesty \cite{harmonjones2019introduction}. The social norms framework highlights normative mechanisms that shape perceptions of peer behavior and expectations, either normalizing or discouraging cheating \cite{lois2021honest}. Finally, self-efficacy theory points to confidence-building mechanisms, such as verbal persuasion or mastery experiences, which enable students to resist dishonest shortcuts when faced with challenging tasks \cite{murdock2006motivational}. This theory-informed, motivation-based intervention design addresses three research questions:

\begin{itemize}
    \item \textbf{RQ1:} a) Do concept-based interventions reduce cheating behavior? b) Do effects vary by concepts?
    
    \item \textbf{RQ2:} a) Do concept-based interventions affect performance and experience across integrity groups? b) Do these effects differ by concept?
    
    \item \textbf{RQ3:} How do concept-based interventions influence cheating behavior, performance, and experience across integrity groups through participants' self-rated psychological states or mechanisms?
\end{itemize}

To answer these research questions, this study employs a mixed-methods design. We first identified 15 psychological concepts across four theoretical frameworks, and an expert workshop translated each into three targeted messages, producing a total of 45 motivational messages. \autoref{tbl:concepts} summarizes these concepts, their behavioral implications, and example messages. Using a between-subjects design, participants were randomly assigned to receive one of the 45 messages or no message while completing a time-limited, reward-based anagram challenge remotely. The interventions, designed to promote honest performance and positive assessment experiences, were evaluated on cheating behavior, task performance, and assessment experience, as well as participants' perceptions of message impact and the underlying psychological mechanisms specific to each theory.

\begin{table*}[!ht]
\centering
\caption{Theoretical concepts, underlying internal psychological states or ``mechanisms'', and message examples by theoretical framework}
\label{tbl:concepts}
\small
\begin{tabular}{p{2.2cm}p{2.2cm}p{2.5cm}p{4.5cm}p{3.5cm}}
\toprule
\textbf{Framework} & \textbf{Concept} & \textbf{Definition} & \textbf{Example Message} & \textbf{Psychological Mechanisms} \\
\midrule
\multirow{3}{2.2cm}{Self-determination} 
& Autonomy & Personal control and volition & ``It's up to you how you tackle these challenges. You have the choice...'' & \multirow{3}{3.5cm}{Autonomy satisfaction/frustration, Competence satisfaction/frustration, Relatedness satisfaction/frustration} \\
\cmidrule{2-4}
& Competence & Feelings of effectiveness & ``When you're working on something and it finally clicks? It's like unlocking a new level...'' & \\
\cmidrule{2-4}
& Relatedness & Connection with others & ``Can you think about all the people impacted by how you behave? Your work is connected...'' & \\
\midrule
\multirow{4}{2.2cm}{Cognitive dissonance}
& Self-concept & Core identity and values & ``What makes you, you? When you make choices that honor those qualities...'' & \multirow{4}{3.5cm}{Cognitive discomfort, Moral disengagement} \\
\cmidrule{2-4}
& Cognitive inconsistency & Conflicting thoughts/behaviors & ``What's your gut telling you? Do you sense any contradictions...'' & \\
\cmidrule{2-4}
& Dissonance arousal & Discomfort from misalignment & ``When you don't quite give it your all? You'll feel it---that uneasy feeling...'' & \\
\cmidrule{2-4}
& Dissonance reduction & Aligning actions with values & ``Can you align your actions and values to feel more in sync?'' & \\
\midrule
\multirow{4}{2.2cm}{Self-efficacy}
& Performance accomplishments & Past successes & ``What challenge have you overcome? It's proof you've got the skills...'' & \multirow{4}{3.5cm}{Perceived performance accomplishment, vicarious experience, verbal persuasion, emotional arousal} \\
\cmidrule{2-4}
& Vicarious experience & Learning from others & ``Can you recall when someone overcame a similar hurdle? It's a confidence booster...'' & \\
\cmidrule{2-4}
& Verbal persuasion & Encouragement & ``You've got the skills, hustle, and determination it takes...'' & \\
\cmidrule{2-4}
& Emotional arousal & Positive emotions & ``What emotions do you experience when tackling challenges? Your emotions are fuel...'' & \\
\midrule
\multirow{4}{2.2cm}{Social norms}
& Descriptive norms & Typical behavior & ``Can you think of someone tackling a project with honesty? We often look to others...'' & \multirow{4}{3.5cm}{Perceived descriptive norms, injunctive norms, social sanctions, reference group identification} \\
\cmidrule{2-4}
& Injunctive norms & Social expectations & ``Are you thinking about how others view your actions...'' & \\
\cmidrule{2-4}
& Social sanctions & Consequences & ``What's at stake? You're building a reputation, not just something...'' & \\
\cmidrule{2-4}
& Reference group identification & Group membership & ``Are you aware you're part of a community dedicated to best abilities...'' & \\
\bottomrule
\end{tabular}
\end{table*}\label{sec:intro}

\section{Related Work}
Exam cheating, one of the academic dishonesty practices, has become a critical concern in educational institutions worldwide. Recent studies reveal the magnitude of this problem, with rates ranging from over 80\% among college graduates to 95\% among high school students admitting to various forms of academic misconduct, e.g., copying from others in exams, plagiarism, consulting unauthorized materials during exams etc. \cite{nora2010motives, yardley2009true}.

\subsection{Key factors influencing cheating behavior}

High-stakes situations, such as university finals, admissions tests, or competitive exams, can create intense pressure that pushes some students toward dishonest choices \cite{fontaine2020exam, kam2018academic}. Past studies observed that students' decisions to cheat are usually shaped by a combination of individual factors, including gender, grades, self-esteem, academic aptitude, perseverance, self-discipline, and moral reasoning, which guides them in distinguishing right from wrong, as well as situational influences, such as peer behavior, prevailing ethical norms, and administrative deterrence \cite{amigud2019246, ghanem2019study, mccabe2001dishonesty}. While moral reasoning can deter dishonest behavior, studies suggest that the perceived likelihood of being caught often has a stronger impact than the severity of potential punishment \cite{freiburger2017cheating, miller2011reasons}.

\subsubsection{Cheating behavior types}

Building on these individual and situational factors, past studies have also recognized that cheating behavior itself may exist along a spectrum, ranging from complete honesty to consistent dishonesty. Studies on academic dishonesty, especially exam cheating, indicate that it is not a simple yes-or-no behavior. Some students remain entirely honest, others cheat extensively, and many engage partially, doing only what they feel they can justify \cite{fischbacher2013lies, mazar2006dishonesty}. This spectrum reflects how frequently students engage in cheating across assessment items, whether on none, some, or all questions, rather than the degree of assistance they use on individual items. These patterns correspond not only to behavior but also to psychological differences between students. For example, studies show that partial-cheaters typically balance self-image with material gain, full-cheaters act opportunistically when risks are low, and non-cheaters rely on internalized norms to guide their honesty \cite{pascualezama2020cheaters}. These distinctions have important implications for interventions. Moral reminders or self-concept nudges often reduce partial cheating but have little effect on full-cheaters \cite{mazar2008dishonesty}, whereas norm-based strategies tend to reinforce honesty among honest ones \cite{cialdini2006managing}. Recognizing these differences allows for more nuanced, context-sensitive prevention strategies, an issue of growing importance in unproctored online assessments, where technology increases opportunities for dishonesty \cite{king2009online}.

\subsection{Message-based interventions in unproctored remote assessments}

Unproctored remote assessments pose unique challenges for assessment integrity because test-takers are not directly monitored. These assessments rely on students' self-regulation and understanding of rules, making interventions that foster motivation toward honest participation particularly important \cite{kangwa2024selfdoubt}. Institutions often prefer unproctored assessments for their flexibility, cost-effectiveness, and to avoid privacy concerns associated with remote proctoring \cite{rios2017online}. However, these exams require careful design to discourage exam cheating. Conventional measures, such as time limits, randomized questions, restrictions on backtracking, and critical-thinking tasks, can reduce cheating opportunities, yet detection remains difficult, and some students still use unauthorized help \cite{chen2018how, harper2006high, schultz2022ok}. Detecting dishonest behavior has become even more challenging with the rise of AI tools, which can generate original, coherent, and contextually appropriate responses that evade traditional methods like answer matching, plagiarism checks, or similarity-based detection systems \cite{cotton2024chatting, kazley2025use}.

Among the most widely studied approaches to address cheating in unproctored remote assessments are textual interventions, particularly message-based strategies. Honor codes, for instance, shift responsibility for integrity to students and foster personal accountability, increasing understanding of cheating and reducing dishonest behavior when reinforced \cite{mccabe1999academic, tatum2022honor, tatum2017honor}. Explicit penalty warnings and fake surveillance messages can further reduce cheating, especially when combined with honor code reminders \cite{bing2012experimental, pleasants2022cheating}. Recent comparisons indicate that honor code reminders are often the most effective, as they activate moral reasoning, whereas other interventions mainly prompt students to adjust strategies without fully preventing dishonesty \cite{mukherjee2023effects}. Other promising approaches include social norm nudges that highlight honesty as typical, autonomy-supportive messages encouraging ethical behavior, and positive reinforcement that rewards integrity \cite{henderson2024auditing, miller1975attribution}.

\subsubsection{Shortcomings of current message-based interventions}

Despite these advances, most interventions remain ad hoc, intuitively designed, relying on surface-level cues such as warnings or honor code reminders without systematically targeting the psychological drivers of cheating. This creates three critical problems that justify the need for theory-driven interventions.

\textit{First}, this intuitive design lacks mechanistic clarity. Current interventions cannot explain how they work. For example, when an honor code reminder reduces cheating, we cannot determine whether it activates moral values, increases fear of consequences, shapes peer behavior perceptions, or builds test-taker's confidence, or some combination thereof. Without understanding these underlying mechanisms, practitioners cannot predict which messages will succeed or fail, limiting their ability to design effective interventions. \textit{Second}, such design has unintended consequences. Poorly designed interventions risk creating side effects that undermine assessment validity itself. For instance, fear-based messages may heighten test anxiety or impair performance among honest test-takers \cite{chin2017influence, mukherjee2023effects, wuthisatian2020student}. \textit{Third}, such practice has limited generalizability. Ad hoc approaches offer no principled basis for adapting interventions across contexts. What works in one assessment setting may fail in another for untested reasons, making it difficult to justify, replicate, or scale these interventions systematically across educational institutions or assessment types.

Research suggests that theory-driven interventions can address these gaps by explicitly identifying the underlying psychological processes, such as motivation, self-concept, perception of social norms, and self-efficacy beliefs, that messages are designed to influence \cite{murdock2006motivational, rimal2005how, stephens2017how}.

\subsection{Theoretical framework for cheating intervention design}

So far, we have reviewed the challenges of unproctored assessments and the limitations of conventional and textual anti-cheating interventions, highlighting the need for theoretical grounding in message design. For that purpose, we identified four psychological theoretical frameworks to guide interventions for unproctored assessments, as each addresses a distinct aspect of cheating behavior. \textit{Self-determination theory} highlights how satisfying or frustrating test-takers' needs for autonomy, competence, and relatedness influences motivation and susceptibility to cheating. \textit{Cognitive dissonance theory} explains how test-takers reconcile conflicts between their actions and self-concept, revealing opportunities to encourage ethical alignment. \textit{Social norms} framework focuses on how perceptions of peer behavior and expectations could shape dishonest behavior. \textit{Self-efficacy theory} emphasizes confidence and mastery as key factors in reducing reliance on cheating. Together, these theories cover motivation, moral reasoning, social influence, and capability beliefs, forming a comprehensive basis for designing multi-faceted interventions. Below we discuss these four identified theoretical lenses, to examine how their core concepts can influence students' choices to cheat, as well as their performance and experience, also through influencing underlying psychological processes.

\subsubsection{Self-determination to understand cheating behavior}

\paragraph{\textbf{Foundations and key concepts of self-determination theory}}

Self-determination theory (SDT) reconceptualizes motivation along a continuum from controlled to autonomous regulation, challenging traditional views of motivation as simply high or low \cite{ryan2000self}. Controlled motivation includes external regulation (acting for rewards/punishments) and introjected regulation (acting from guilt or ego), while autonomous motivation encompasses identified regulation (personal importance) and intrinsic motivation (inherent enjoyment).

SDT's core proposition is that three basic psychological needs drive human behavior: autonomy (acting from personal volition rather than external control), competence (feeling effective and capable), and relatedness (experiencing connection and belonging). When these needs are satisfied, individuals experience autonomous motivation and align behavior with personal values. When these needs are not met, they resort to controlled behaviors that may justify dishonest means to achieve desired ends \cite{reeve2012self, vansteenkiste2010autonomous}.

\paragraph{\textbf{Application of self-determination in cheating research}}

Empirical evidence shows that autonomous motivation reduces cheating, whereas controlled motivation often increases it. Students who find their education meaningful and engaging are less likely to cheat, while those focused on external rewards (e.g., grades, job prospects, or social approval) are more prone to see cheating as a shortcut when effort does not guarantee success \cite{anderman1998motivation, davy2007examination, murdock2001predictors, newstead1996individual, pulfrey2013why}. Basic needs such as autonomy, competence, and relatedness, offer distinct pathways shaping dishonesty and provide a framework for how environments can foster or prevent misconduct.

Autonomy matters because controlling environments that restrict choice and impose pressures increase cheating, while autonomy-supportive classrooms that respect student perspectives reduce it \cite{bureau2014parental}. Competence also guides behavior: students with high self-efficacy and clear expectations cheat less, whereas low confidence or unclear instruction raises cheating rates \cite{finn2004academic, murdock2004effects}. Relatedness further protects against misconduct, as students who feel supported and cared for by teachers are less likely to cheat \cite{calabrese1990relationship, murdock2001predictors}. Together, these needs interact to promote academic integrity within ethical learning environments.

Yet SDT alone does not capture the full complexity of cheating. Motivation interacts with cognitive and social processes: students high in self-enhancement values may misremember mastery-oriented instructions as competitive, reshaping reality to justify dishonesty \cite{pulfrey2019under}. Competence links closely with self-efficacy, where mastery experiences shape confidence \cite{bandura1997self, usher2009sources}, but students can inflate self-efficacy through cheating and self-deception \cite{wedge2012roles}. These findings show that cognitive dissonance, social norms, and self-efficacy provide complementary lenses, helping explain how students navigate internal conflicts when tempted to cheat.

\subsubsection{Cognitive dissonance in cheating understanding}

\paragraph{\textbf{Foundations and key concepts of cognitive dissonance theory}}

Cognitive dissonance theory, developed by Festinger \cite{festinger1957theory}, explains the psychological mechanisms underlying academic dishonesty. It proposes that individuals experience discomfort when their behavior conflicts with beliefs, attitudes, or values \cite{festinger1957theory, harmonjones2019introduction}. In cheating, this tension arises because most students believe dishonesty is wrong yet still may occasionally behave dishonestly \cite{stephens2017how}. Festinger \cite{festinger1962cognitive} outlined two core principles: dissonance creates discomfort that motivates reduction, and people tend to avoid situations or information that might increase discomfort. Crucially, cognitive inconsistency refers to the contradiction between cognitions, while dissonance denotes the resulting psychological discomfort, which manifests as tension, aversive arousal, or negative affect \cite{harmonjones2019introduction}.

The action-based model suggests dissonance emerges when conflicting cognitions affect potential actions, signaling a need to restore consonance for effective behavior \cite{harmonjones2019introduction}. Students reduce dissonance through various strategies: changing attitudes, adjusting behavior, or reinterpreting the importance of conflicting cognitions \cite{festinger1957theory}. They may also avoid information that could heighten dissonance, such as skipping honor codes \cite{stephens2017how}, and rationalize their actions to maintain a positive self-concept despite ethical conflict.

\paragraph{\textbf{Application of cognitive dissonance in cheating research}}

Cognitive dissonance theory has been widely applied to understand how students manage the conflict between moral beliefs and dishonest behavior. To reduce dissonance, students often rationalize or morally disengage, temporarily justifying cheating while preserving a positive self-concept \cite{bandura1990mechanisms, stephens2017how, sykes1957techniques}. Social influences can amplify this process: perceiving that ``everyone else is doing it'' makes cheating easier to reconcile with personal values, even when this perception is inaccurate (pluralistic ignorance; \cite{zhao2019effects}).

Empirical evidence supports these mechanisms. Interventions such as reading or signing honor codes and completing plagiarism tutorials reduce cheating by raising ethical awareness and limiting rationalizations \cite{shu2011dishonest}. Yet some approaches, like induced-hypocrisy, a technique that reminds people of past inconsistencies between their values and actions to motivate behavior change, have had mixed successes, illustrating the complexity of applying such methods in real-world contexts \cite{vinski2009study}.

Cognitive dissonance often intersects with other frameworks. Moral disengagement allows students to temporarily set aside moral standards without changing underlying beliefs \cite{bandura1990mechanisms, stephens2017how}, while perceived descriptive norms provide external justification, mediating the link between social expectations and dishonest intentions \cite{zhao2019effects}. Causal attributions further help students manage responsibility and reduce dissonance. Attribution theory identifies three key dimensions: locus (internal versus external causes), stability (temporary versus permanent factors), and controllability (whether the cause can be influenced by the individual) \cite{weiner1995judgments}. For example, attributing failure to lack of effort (internal, unstable, controllable) creates less dissonance than attributing it to lack of ability (internal, stable, uncontrollable), potentially reducing motivation to cheat. Complementing SDT, cognitive dissonance clarifies how students navigate conflicts between intrinsic values, such as integrity, and external pressures, sometimes suppressing autonomous motivation in favor of controlled motivation while using rationalizations. Integrating these frameworks illuminates the complex psychological processes that allow students to reconcile moral beliefs with cheating behavior.

\subsubsection{Social norms approach in cheating understanding}

\paragraph{\textbf{Foundations and key concepts of social norms}}

The social norms approach encompasses multiple frameworks examining how perceived and actual norms shape behavior \cite{lapinski2005explication}. Its roots lie in early social psychology, with Sherif \cite{sherif1936psychology} defining norms as the standardized customs, values, and rules that guide conduct through social interaction. Norms operate at two levels: collectively, as prevailing codes that guide group behavior, and psychologically, as individuals' perceptions or misperceptions of those codes \cite{lapinski2005explication}. These perceptions often drive behavior more than actual norms, making them central to cheating research \cite{stephens2017how}.

A key distinction exists between descriptive norms, which describe typical behavior, and injunctive norms, which reflect social approval or disapproval \cite{rimal2005how}. Pluralistic ignorance occurs when students mistakenly believe peers accept a norm while most actually reject it, such as assuming that most peers cheat, which can normalize dishonest behavior and lower resistance to cheating oneself \cite{prentice1993pluralistic}.

Several frameworks detail how norms influence behavior. The Theory of Normative Social Behavior (TNSB) highlights that descriptive norms shape intentions through group identity, injunctive norms, and outcome expectations \cite{rimal2005how}. The Focus Theory of Normative Conduct shows that norms guide behavior only when salient, with descriptive norms signaling what works and injunctive norms conveying social rewards or sanctions \cite{cialdini1990focus}. The Theory of Reasoned Action \cite{fishbein1975belief} and the Theory of Planned Behavior \cite{ajzen1991theory} integrate subjective norms as social pressures motivating intentions, emphasizing injunctive influences \cite{rimal2005how}.

\paragraph{\textbf{Applications of social norms in cheating research}}

Research consistently shows that perceived peer norms strongly influence academic dishonesty. Peer norms are among the strongest predictors of cheating, with students reporting less misconduct when they perceive peers disapprove and more when they believe peers cheat \cite{mccabe1997individual}. Observations of peer cheating are common. Indeed, 97\% of students reported witnessing peers cheating, which leads many to assume that most peers engage in dishonest behaviors \cite{henningsen2020cheating}. Descriptive norms interact with group identity, injunctive norms, and outcome expectations to shape cheating intentions.

Pluralistic ignorance amplifies this effect: students often misperceive peers' behavior, justifying their own cheating as normal and creating a self-reinforcing cycle \cite{henningsen2020cheating}. Experimental studies confirm causal effects: students are more likely to cheat after observing an in-group member cheat, whereas out-group cheating has less influence \cite{gino2009contagion}, and awareness of others' dishonesty increases individual misconduct through conformity \cite{fosgaard2013separating}.

Interventions using social norms yield mixed results. While normative feedback reduces behaviors like alcohol use \cite{lapinski2005explication}, emphasizing descriptive norms in contexts where cheating is common can backfire and normalize dishonesty instead of deterring it \cite{cialdini1990focus}. For example, informing students that ``many of your peers use unauthorized resources during online exams'' may inadvertently signal that cheating is widespread and acceptable, thereby increasing rather than decreasing dishonest behavior. Social norms also intersect with cognitive and motivational processes: believing ``everyone else is cheating'' can reduce moral discomfort via cognitive dissonance \cite{stephens2017how}, and moral disengagement mediates the relationship between perceived norms and corrupt intentions \cite{zhao2019effects}. From a self-determination perspective, norms can foster autonomous motivation when supporting integrity or impose external pressures when favoring cheating \cite{pulfrey2019under}. Norms can also shape attributions, allowing students to blame situational factors while protecting self-concept.

Integrating social norms with motivational, moral, and attributional frameworks clarifies how perceived peer behavior can both explain why students cheat despite knowing it is wrong and illuminate how social environments promote or undermine academic integrity.

\subsubsection{Self efficacy theory in cheating understanding}

\paragraph{\textbf{Foundations and key concepts of self-efficacy}}

Self-efficacy, introduced by Bandura \cite{bandura1977self}, is a central concept in social cognitive theory \cite{bandura1986social}. It provides a lens to understand how perceived competence shapes academic dishonesty. It is defined as ``beliefs in one's capabilities to organize and execute the courses of action required to produce given attainments'' \cite{bandura1997self}. Related but distinct are outcome expectations, beliefs about likely results, which may align or diverge from self-efficacy. For example, a student may feel capable in mathematics but avoid an advanced course due to expected unfair grading \cite{usher2009sources}.

Self-efficacy develops from four main sources \cite{bandura1977self}. The strongest, performance accomplishments or mastery experiences, builds confidence through success and undermines it through repeated failure. Students who succeed honestly are more likely to trust their ability to achieve without cheating, whereas those attributing failure to lack of ability may turn to dishonesty. Vicarious experience, gained by observing similar peers succeed, reinforces beliefs in one's own competence. Verbal persuasion from teachers, parents, or peers strengthens confidence, though less powerfully than mastery or vicarious experiences. Finally, physiological and emotional states influence self-efficacy: anxiety or stress can signal unpreparedness and increase temptation to cheat, whereas positive emotions signal readiness and support honest performance \cite{usher2009sources}.

A distinction exists between task-specific and general self-efficacy. Task-specific self-efficacy is tied to particular domains, such as mathematics, while general self-efficacy reflects broader confidence across contexts \cite{bandura1997self, chen2001validation, wedge2012roles}. Task-specific self-efficacy is tied to particular domains (e.g., ``I am confident in mathematics''), while general self-efficacy reflects a cross-domain belief in one's overall capability to handle challenges (e.g., ``I can generally succeed at new things I try''). General self-efficacy develops through accumulating successes across different types of tasks and contexts and becomes particularly important for predicting behavior in novel situations where domain-specific experience is lacking.

\paragraph{\textbf{Applications of self efficacy in cheating research}}

Because self-efficacy varies by domain, its effects on academic dishonesty are best studied at the task-specific level \cite{pajares1996self}. Research shows that students with high task-specific self-efficacy (i.e., confidence in completing assignments or exams honestly) are consistently less likely to cheat, while low-efficacy students may rely on dishonest strategies to compensate for perceived inability \cite{finn2004academic, murdock2001predictors}.

Bandura's four sources of self-efficacy shape cheating differently. Performance accomplishments or mastery experiences reduce dishonesty by reinforcing confidence in legitimate strategies, whereas attributing failure to lack of ability increases vulnerability \cite{bandura1997self, lopez1997role}. Vicarious experiences influence cheating through peer comparison: observing similar classmates succeed honestly strengthens confidence and reduces temptation \cite{schunk1987peer, usher2009sources}. Verbal persuasion from teachers or mentors supports honest performance, though its effect is smaller than mastery or vicarious experiences \cite{usher2009sources, nora2010motives}. Physiological and emotional states also matter: anxiety or fear of failure signals low efficacy and increases cheating risk, while positive moods and manageable arousal support integrity \cite{anderman1998motivation, bandura1997self, calabrese1990relationship, malinowski1985moral}.

Self-efficacy interacts with other psychological processes. Low efficacy can threaten competence needs in self-determination theory, promoting externally regulated behaviors like cheating, whereas high efficacy fosters autonomous motivation and reduces dishonesty \cite{kanatmaymon2015role, loi2021when}. It also intersects with cognitive dissonance: students with moderate or high efficacy may rationalize cheating to resolve dissonance, while very low-efficacy students experience less conflict \cite{wedge2012roles}. Social norms amplify these effects, as low-efficacy students are more influenced by peers' dishonest behavior, whereas confident students resist pressure \cite{nora2010motives}. Attribution processes further modulate cheating risk, with controllable interpretations of failure preserving efficacy and stable internal attributions undermining it \cite{bandura1997self}.

Overall, according to this theory, self-efficacy shapes cheating through a complex interplay of beliefs, motivation, social context, and cognitive processes, offering critical insights for interventions that strengthen competence, reduce reliance on dishonesty, and promote academic integrity.

\subsection{Game-based paradigms for remote assessment research}

Understanding how students behave in controlled experimental settings is crucial for designing theory-informed interventions in unproctored remote assessments. After reviewing preventive practices, the need for theory-driven message design, and the foundations of four popular behavioral theories and their impact on cheating, it is important to examine research tools that capture dishonest behavior reliably.

Experimental research increasingly uses cognitive games as proxies for real-world testing, providing controlled environments to study cheating while maintaining ecological validity. Common paradigms include matrix tasks \cite{mazar2008dishonesty}, die-rolling experiments \cite{fischbacher2013lies}, and mind games \cite{jiang2013cheating}, which simulate the cognitive engagement, time pressure, and reward structures of high-stakes academic assessments. Our study builds on this tradition using anagram challenge, which requires pattern recognition, vocabulary recall, and strategic thinking, skills analogous to those demanded in academic assessments, while remaining accessible across a wide range of educational backgrounds \cite{abeler2019preferences}.

The time-constrained, reward-based structure of the anagram challenge mirrors remote exam conditions, where students must exhibit competence under pressure and face performance-based consequences. Online availability of solutions creates realistic cheating opportunities, enabling unobtrusive detection through behavioral monitoring rather than invasive surveillance. Meta-analytic evidence supports this approach, showing that laboratory-based cognitive tasks validly measure dishonest tendencies that generalize to real-world contexts \cite{gerlach2019truth}. Consequently, our anagram paradigm provides a meaningful and ethically sound proxy for studying integrity interventions in remote assessment environments.\label{sec:literature}

\section{Study Design}
Unlike previous studies that often designed messages incorporating multiple components of a single theory, our approach isolates concepts at the individual level by designing separate messages for each concept. This allows us to examine, with greater precision, how self-rating of specific psychological mechanisms from four established behavioral theories shape honest and motivated behavior in remote assessments. We conducted a two-part approach.

In Part 1, we carried out message design based on behavioral concepts. Post literature review, we selected 15 psychological concepts from four established behavioral theories to create motivational messages. These concepts, detailed in \autoref{tbl:concepts}, span self-determination theory (autonomy, competence, relatedness), cognitive dissonance theory (self-concept, cognitive inconsistency, dissonance arousal, dissonance reduction), self-efficacy theory (performance accomplishments, vicarious experience, verbal persuasion, emotional arousal), and social norms framework (descriptive norms, injunctive norms, social sanctions, reference group identification). Through an online expert workshop, we created three motivational message variations per concept, resulting in 45 theory-informed interventions.

In part 2, i.e. the evaluation phase, we designed a remote website featuring a reward-based, time-limited anagram task to evaluate these concepts. Participants were randomly assigned either to one of the concept-specific messages or to a control condition without any message, using a between-subjects design. This setup allowed us to assess how the interventions influenced cheating behavior, task performance, and assessment experience, as well as the self-rated internal psychological states or mechanisms (e.g., need satisfaction or frustration, normative perceptions, cognitive discomfort) driving these effects.\label{sec:design}

\section{Methodology}
\subsection{Experimental design}

\subsubsection{Design of a website for anagram challenges}

\begin{figure}[!ht]
\centering
\includegraphics[width=0.7\textwidth]{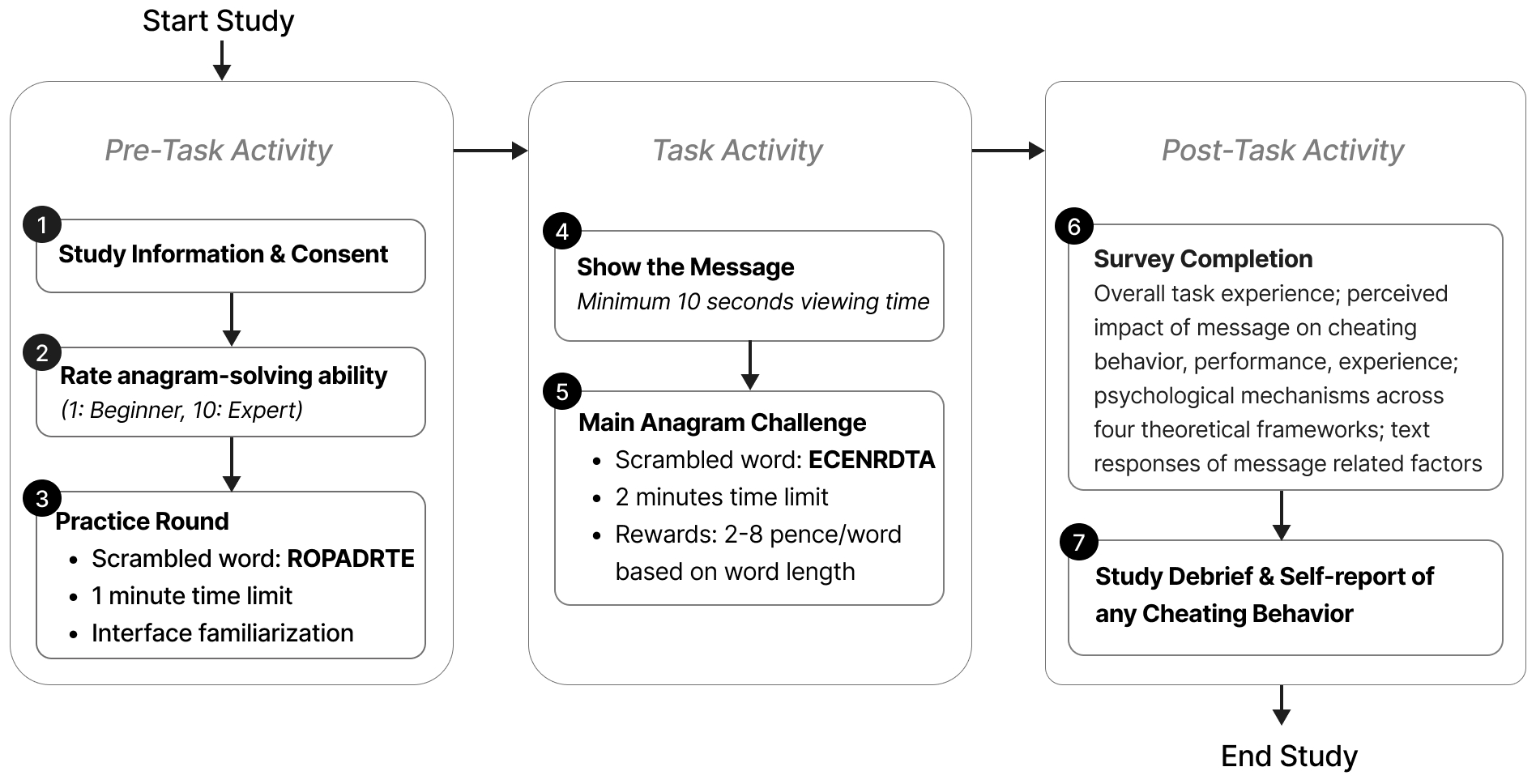}
\caption{Sequential study flow for evaluating concept-based messaging interventions in a remote unproctored anagram challenge. Participants rated their ability to solve anagram (2), practiced the task (3), viewed a theory-based message to motivate honest performance (4), completed two 2-minute rounds (5) where longer and more numerous words earned higher rewards, then filled out a post-task survey (6) and were debriefed, including disclosure of self-reported cheating (7).}
\label{fig:study_flow}
\end{figure}

\begin{figure}[!ht]
\centering
\includegraphics[width=0.85\textwidth]{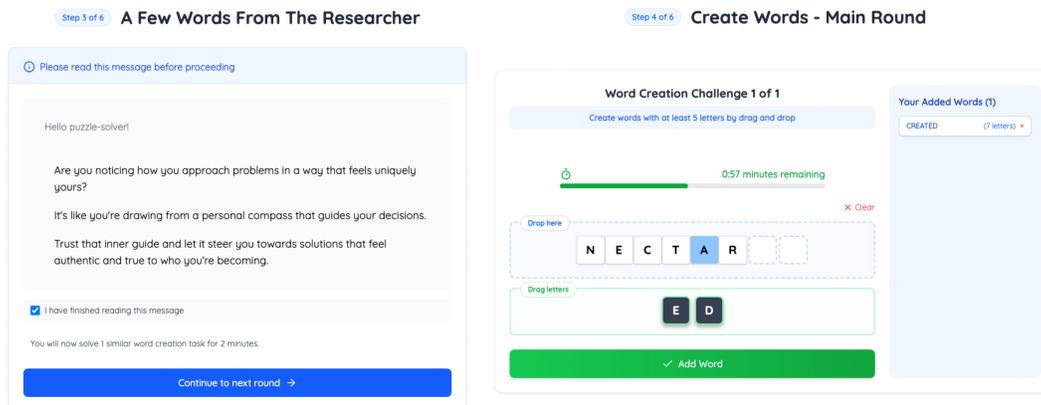}
\caption{Screenshots of the online study interface. Left: Message interventions displayed for a mandatory 10-second period. Right: Participants created words using drag-and-drop interactions.}
\label{fig:interface}
\end{figure}

We developed a web-based interface to examine how different messaging approaches influence honest and motivated performance. \autoref{fig:study_flow} shows the overall study flow. The interface provided an engaging experience while enabling detailed tracking of interactions and performance. All task activities were conducted through this website; participants completed post-task questionnaires on a separate survey portal. \autoref{fig:interface} presents screenshots of the participant view.

\textbf{Selection of anagram challenge as a cognitive task:} The anagram task served as an effective proxy for remote cognitive assessments while aligning with research aims. Participants generated valid words from scrambled letters which is a cognitively demanding activity involving pattern recognition, vocabulary retrieval, and strategic thinking, skills similar to academic problem-solving. Anagram paradigms are widely used in cognitive psychology to manipulate psycholinguistic variables such as word frequency and letter combinations \cite{valerjev2020impact}. The task's accessibility across educational levels enabled broader participation and robust statistical inference, balancing experimental control and ecological validity in remote settings.

\textbf{Study flow and structure:} After reviewing instructions, data privacy information, and giving consent (Phase 1), participants were verified to be using a laptop or desktop (required for the drag-and-drop task). In Phase 2, they completed a one-minute practice round with scrambled letters (``ROPADRTE''). In Phase 3, a concept-specific message appeared for a mandatory 10 seconds based on prior findings that online participants often skip instructions when eager to proceed \cite{arechar2018conducting}. In Phase 4, participants undertook the main 2-minute reward-based challenge using new letters (``ECENRDTA''). In Phase 5, they completed a post-task survey on task experience, message impact, their psychological experiences related to the psychological processes (e.g., feelings of autonomy, competence, relatedness, cognitive discomfort, perceived norms, self-efficacy beliefs), and demographics. Finally, Phase 6 involved debriefing, revealing the study's true purpose and asking about use of external resources.

\textbf{Performance incentives in anagram challenge:} The 2-minute time limit given in the main challenge round was to simulate real-world assessment pressure. The 8-letter anagram ``ECENRDTA'' was selected through pilot testing for its moderate difficulty. Participants were required to generate valid words of at least five letters. A performance-based reward structure incentivized both quantity and quality: 5-letter words earned 2 pence, 6-letter words 4 pence, 7-letter words 6 pence, and 8-letter words 8 pence, up to a maximum of 25 pence per round in addition to a £2.40 base payment. Although all participants ultimately received the maximum bonus, they were told bonuses depended on performance to simulate real-world motivation. This mild deception, disclosed during debriefing, simulated real-world motivation without penalizing honest participants.

\textbf{Behavioral monitoring and cheating detection:} To detect potential cheating during the anagram challenge, several unobtrusive monitoring mechanisms were implemented. The system continuously tracked when participants navigated away from and returned to the browser tab containing the challenge, flagging cases where a surge in word submissions occurred immediately afterward, potentially indicating the use of external word-finding tools. It also logged mouse activity patterns, identifying periods of inactivity followed by rapid, accurate submissions as possible evidence of external assistance (e.g., mobile devices or online anagram solvers). Following the task, during the debriefing phase, participants were also explicitly asked to disclose any use of external resources, with clear assurance that their payment would not be affected. However, only 4.2\% of the 473 participants behaviorally flagged as cheaters admitted doing so, underscoring self-report limitations and the need for behavioral tracking as the primary detection method. The full multi-phase cheating detection algorithm is described in the Measurements section below.

\subsubsection{Designing messages aligned with relevant theoretical concepts}

\begin{figure}[!ht]
\centering
\includegraphics[width=0.7\textwidth]{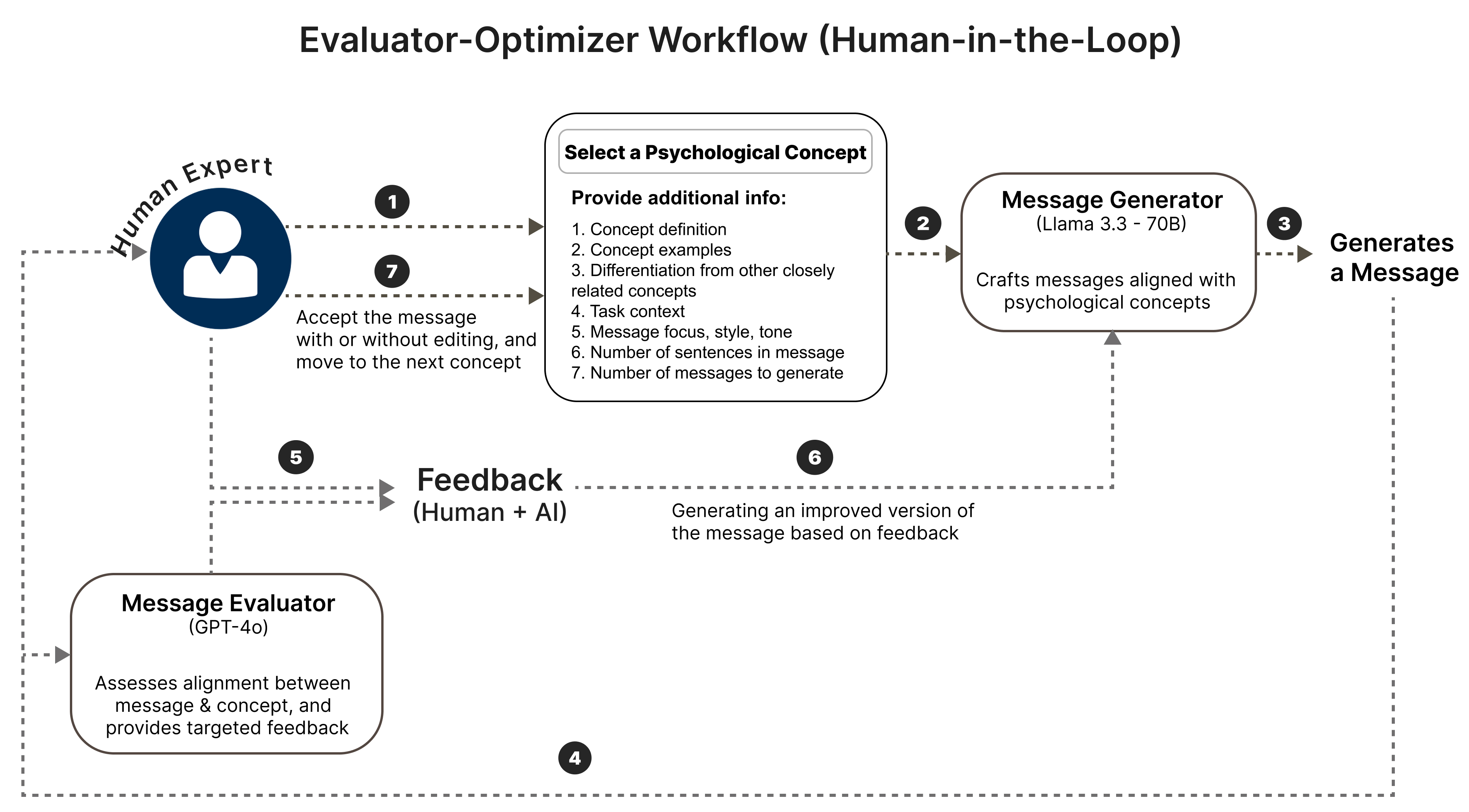}
\caption{Methodology for operationalizing concepts into messages. Experts selected a concept via an in-house web app and generated messages with assistance from a tailored LLM primed with concept definitions and examples. Messages were iteratively refined via an evaluator-optimizer workflow \cite{madaan2023self}, with expert feedback incorporated into subsequent LLM-assisted revisions.}
\label{fig:message_methodology}
\end{figure}

Developing psychological concepts-based messages requires bridging theoretical concepts and practical communication. Although concept-based message design may seem straightforward, ensuring both effectiveness and theoretical alignment is challenging \cite{okeefe2012psychological}. To address this, we implemented a two-phase process: Phase 1 involved AI-assisted message generation under expert supervision, and Phase 2 entailed expert-only evaluation for effectiveness and consistency (\autoref{fig:message_methodology}).

Five researchers participated as experts in a two-week remote workshop. Recruited from the first author's professional network, the group included two co-authors and three colleagues. All held advanced degrees (master's or PhD) in behavioral psychology, cognitive science, human-computer interaction, or related fields, and possessed relevant research experience and familiarity with the behavioral theoretical frameworks under study (see \autoref{tbl:concepts}). This ensured sufficient theoretical expertise to assess concept alignment and contribute to theory-informed message development. The workshop, moderated by the first author, was organized into two phases: one week for message generation and another for evaluation, though participants completed tasks at their own pace. The implementation details can be found in the Github repositories shared in the Appendix.

\paragraph{\textbf{Message Generation Framework:}} We developed a web-based framework for generating messages grounded in 15 psychological concepts (see \autoref{tbl:concepts}). To balance theoretical fidelity with linguistic naturalness, we used recent advances in large language models (LLMs), which can generate contextually nuanced, human-like text for psychological research \cite{brown2020language}. Because LLMs may not always mirror human cognition and can show response biases \cite{bender2021dangers}, our approach combined their generative capacity with chain-of-thought prompting and expert oversight. We employed Meta's LLaMA 3.3 (70B), primed with definitions, conceptual distinctions, and examples, to iteratively produce theory-aligned messages.

Each message followed a three-sentence, motivational question--answer format (e.g., ``What happens when you're working on something and don't quite give it your all? You're gonna feel it, right, that uneasy feeling...''). This design promoted reflection while maintaining clarity and avoiding jargon, ensuring accessibility without sacrificing theoretical accuracy. Experts refined messages through an evaluator-optimizer workflow \cite{madaan2023self}. As shown in \autoref{fig:message_methodology}, a GPT-4o model acted as an automated evaluator, working alongside human experts to iteratively improve drafts based on combined feedback. The process continued until each expert validated one message per concept, yielding 75 finalized messages. These were designed to be generic yet theory-consistent, ensuring reusability across studies while remaining relevant to the experimental task.

\paragraph{\textbf{Two-Stage Evaluation Process:}}
To identify the most effective messages among the five generated per concept, we implemented a two-stage expert evaluation in which participants anonymously rated messages created by others. This approach assessed both theoretical alignment and motivational potential. A custom web-based application facilitated the process by randomizing message presentation and recording detailed rating data.

\textit{Stage 1 (Concept identification and alignment):} Experts first viewed others' messages without knowing their intended concept, indicated which concept they believed each represented, and rated alignment with the actual concept on a 10-point scale (0: least to 10: most aligned). This blind evaluation ensured messages conveyed their underlying psychological mechanisms clearly, preserving theoretical precision.

\textit{Stage 2 (Motivational effectiveness):} Experts then reviewed messages grouped by concept, rating how effectively each would encourage honest and motivated effort, and provided qualitative feedback on practical impact. This comparative step identified the most effective ways to translate theory into motivating communication.

\paragraph{\textbf{Final Message Selection:}} For each message, alignment and motivation scores were averaged, giving equal weight to theoretical accuracy and motivational strength. This balanced approach prevented use cases such as highly engaging but misaligned messages from triggering unintended processes. The top three messages per concept were retained, yielding 45 messages across 15 concepts. The most effective messages, reviewed by the first author, featured direct self-reflection prompts, clear examples, natural language, and concise phrasing. The complete set of selected messages is provided in the Appendix.

\subsection{Measurements}

Having established our experimental design and theory-informed message creation process, we next describe the comprehensive measurement approach used to evaluate intervention effectiveness. Our strategy was designed to capture the multi-dimensional impact of concept-based messages across three pillars of assessment validity identified in the introduction: assessment integrity (measured through observed cheating behavior), performance (measured through total scores), and assessment experience (measured through task engagement and satisfaction).

\begin{figure}[!ht]
\centering
\includegraphics[width=0.7\textwidth]{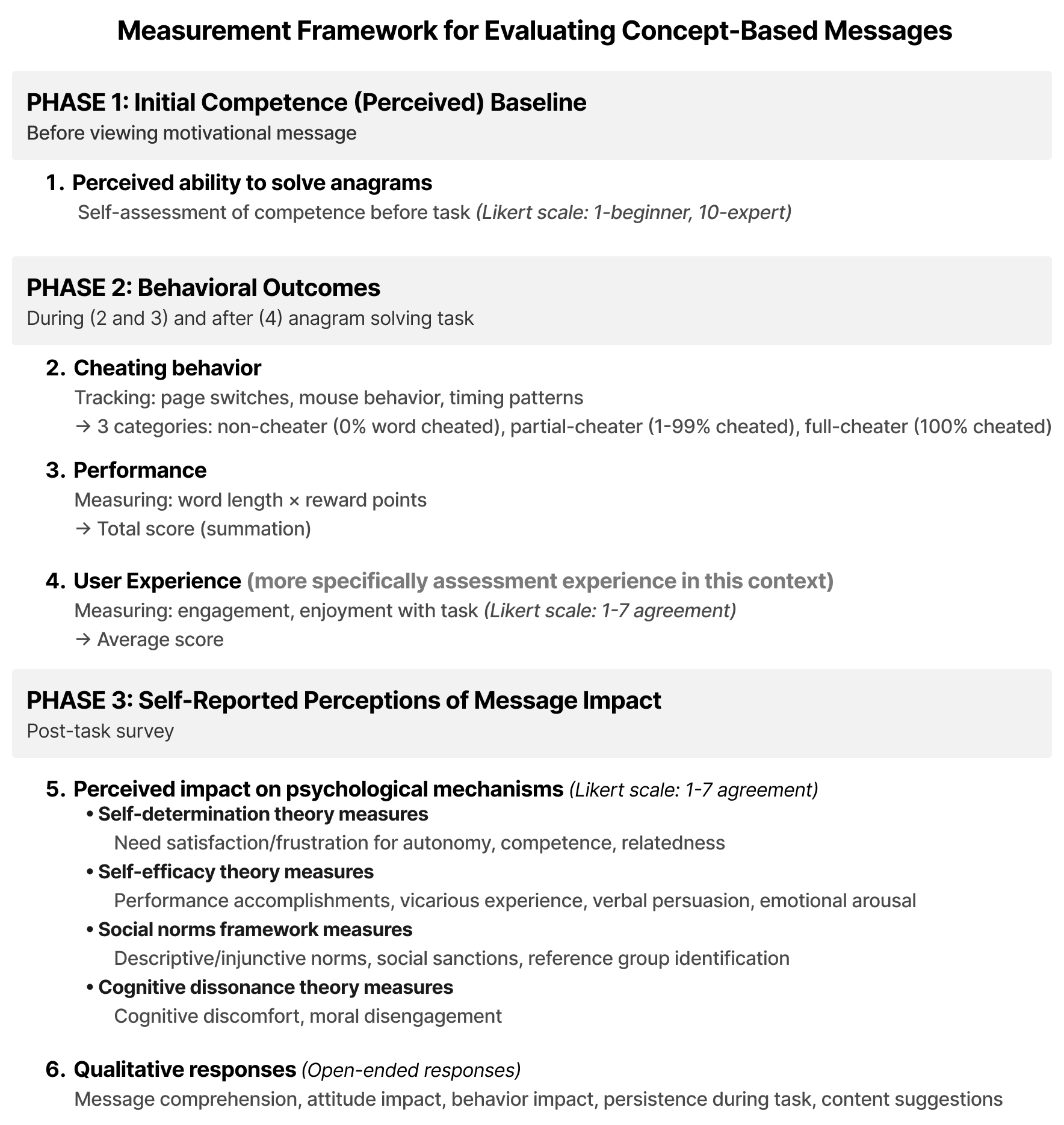}
\caption{Comprehensive measurement framework for evaluating psychological concept-based messages in remote unproctored assessments. Three measurement phases: Phase 1 captures baseline perceived ability before message exposure (1); Phase 2 tracks behavioral outcomes during and after the task (cheating behavior (2), performance (3), experience (4)); Phase 3 assesses self-rated psychological states or mechanisms (5), supplemented by qualitative open-ended questions (6).}
\label{fig:measurement_framework}
\end{figure}

As illustrated in \autoref{fig:measurement_framework}, the measurement framework includes a multi-layered approach that allows us not only to determine whether different psychological concepts may affect cheating, performance and experience, but also to understand the internal mechanisms underlying these effects. Additionally, qualitative responses were collected to gain deeper insights into participants' meaning-making processes and subjective experiences with the interventions.

\subsubsection{Core Outcome Measures}

\paragraph{\textbf{Cheating Behavior}}

\begin{figure}[!ht]
\centering
\includegraphics[width=0.7\textwidth]{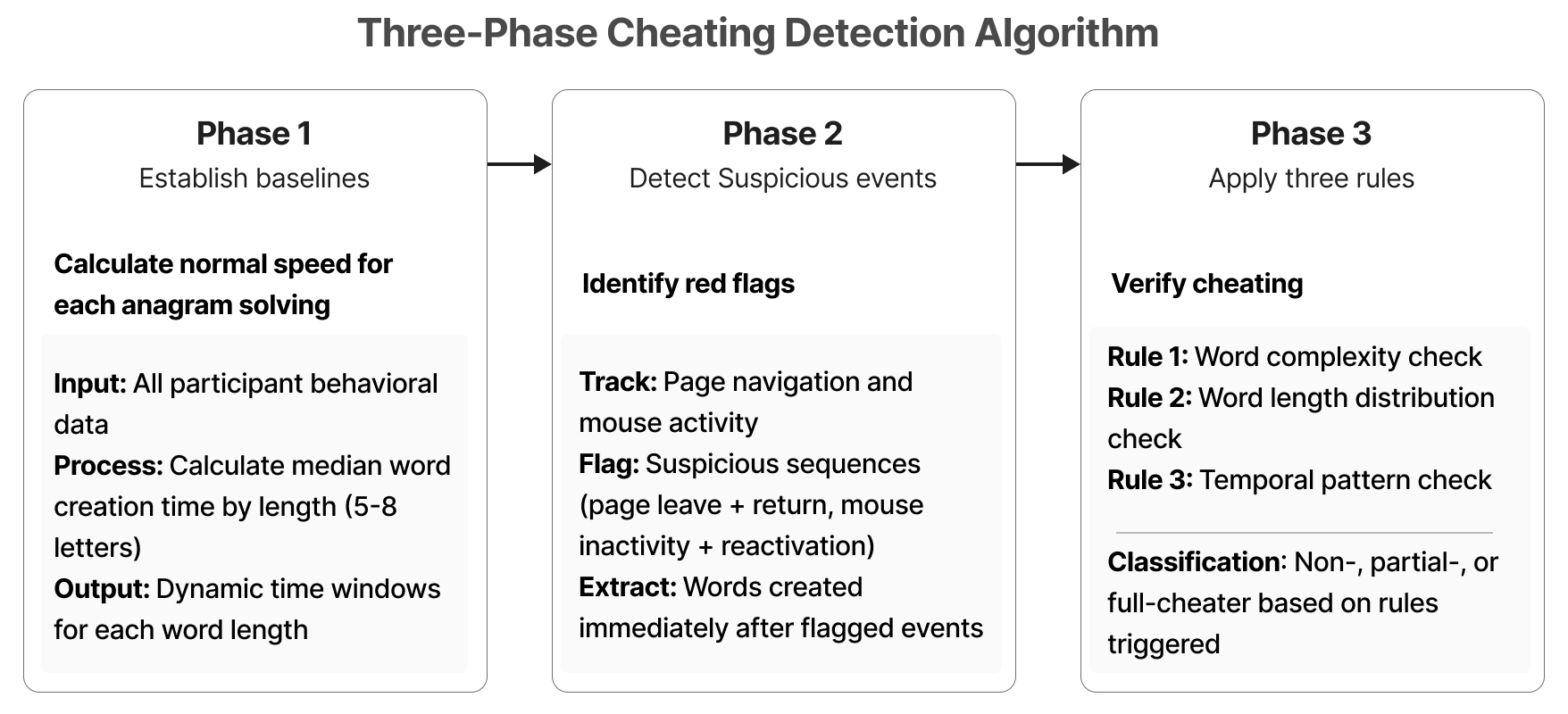}
\caption{Multi-phase cheating detection for remote anagram challenge. The algorithm processes logged user interactions through three main phases: (1) baseline speed establishment, (2) identification of suspicious event sequences, and (3) multi-rule evaluation of flagged behaviors. An additional self-report validation phase was conducted but proved unreliable (only 4.2\% of behaviorally flagged participants admitted cheating), confirming the necessity of behavioral detection methods. Complete algorithmic specifications in Appendix.}
\label{fig:cheating_detection}
\end{figure}

As no standardized method for detecting cheating in similar anagram-game frameworks was identified in the literature, we developed a custom detection algorithm. The development of this algorithm drew upon common behavioral tracking practices and was iteratively refined based on observations from pilot studies. The algorithm prioritized accurate detection of cheating behavior, acknowledging that both false positive and false negative errors are inherent limitations in behavioral observation studies \cite{corrigangibbs2015deterring, pleasants2022cheating}. Detailed procedural steps are illustrated in \autoref{fig:cheating_detection}.

The first phase establishes baseline expectations for word creation speed, ensuring that speed-based flags are meaningful relative to actual participant behavior rather than arbitrary thresholds. Without this step, unusually fast word submissions could not be fairly interpreted, risking both false positives and missed detections.

The algorithm processes behavioral log files automatically recorded during task completion, capturing timestamps for each word submission, word characteristics, and user interaction events (e.g., page navigation, mouse activity). The second phase identifies red-flagged sequences: consecutive word submissions bounded by suspicious behavioral markers such as navigating away from the task or prolonged mouse inactivity. These behavioral markers were selected as potential indicators of external resource consultation. Defining analysis boundaries around these events allows subsequent detection rules to target the specific periods where cheating is most likely, improving precision.

The third phase applies multiple complementary rules to evaluate flagged sequences. Each rule captures a different behavioral dimension: word complexity, shifts in word length distribution, and deviations from expected creation time. We employed multiple criteria because convergent evidence across indicators strengthens our confidence in classifications, though without ground truth data, we cannot quantify true sensitivity and specificity. This conservative approach prioritizes cases where multiple behavioral anomalies co-occur, which we judged more likely to represent genuine cheating than reliance on any single indicator.

Finally, incorporating participant self-reports as a validation step allows cross-checking of behavioral indicators. This mechanism strengthens confidence in the algorithm's accuracy and provides insights into the prevalence of undetected cheating. Overall, the multi-phase, rule-based framework ensures that flagged instances reflect likely cheating, while providing clear, data-driven justifications for each component of the algorithm.

\paragraph{\textbf{Task Performance}}
Task performance was measured as a comprehensive metric capturing participants' total output during the main challenge. The performance score for each participant was calculated as the sum of all rewards assigned to valid words they generated. The performance metric included all valid words, regardless of whether they were identified as cheated. This approach introduces a potential bias: participants who cheated would naturally achieve higher performance scores than their honest ability would yield. However, this bias is acceptable, and even necessary, for our research objectives. Our study examines how psychological interventions influence task engagement and behavior, including the decision to cheat. Separating `honest performance' from `cheating-assisted performance' would obscure the very behavioral patterns we aim to understand. The total performance score thus reflects participants' actual output under different intervention conditions, encompassing both legitimate effort and any external assistance they chose to employ.

\paragraph{\textbf{User experience}}
Participants' experience of the word creation task was measured with a focus on its hedonic dimension. Traditional user experience measures often rely on self-reports, and the Game User Experience Satisfaction Scale (GUESS; \cite{phan2016development}) is a widely used instrument. GUESS includes nine subscales, typically averaged into a composite ``video game satisfaction'' score. For our task, we selected only the Play Engrossment (8 items) and Enjoyment (5 items) subscales, as these can best capture the hedonic aspects relevant to a word creation task; other GUESS subscales (e.g., Narratives, Social Connectivity, Aesthetics) were less applicable to our study. Items were adapted from a gaming to a word creation context and responses were recorded on a 7-point Likert scale from ``strongly disagree'' to ``strongly agree''. Mean scores were computed for each subscale and then averaged to form a single composite representing overall task experience. Internal consistency for the combined scale in our sample was acceptable, supporting the reliability of this measure (Cronbach's $\alpha=0.82$ for engagement and 0.87 for satisfaction). Items for both subscales can be found in the Appendix.

\subsubsection{Self-rated internal psychological states (``mechanisms'') measures}

To examine participants' self-reported psychological states or mechanisms that could reflect the influence of concept-based intervention messages on their behavior, we measured 16 specific mechanisms guided by self-determination, cognitive dissonance, social norms, and self-efficacy frameworks (see \autoref{tbl:concepts}). These measures captured participants' perceptions of how the messages influenced mechanisms rather than directly measuring theoretical mechanisms. The complete set of items used can be found in the Appendix.

\paragraph{\textbf{Psychological need satisfaction and frustration}}
We assessed participants' self-rated satisfaction and frustration of three basic psychological needs using items adapted from the Need Satisfaction and Frustration Scale \cite{longo2016measuring}, modified for the anagram challenge context. Example items for autonomy satisfaction included ``\textit{After reading the message, I felt I had a lot of freedom in deciding how to approach the word creation challenge}''. Need frustration items included ``\textit{After reading the message, I felt incapable of succeeding at the word creation tasks}'' (for competence). Six subscales (i.e., satisfaction and frustration for three needs) were measured on 7-point Likert scales on agreement, with composite scores computed by averaging items within each subscale. Internal consistency was acceptable for both scales (Cronbach's $\alpha = .74$ for need satisfaction and .78 for need frustration).

\paragraph{\textbf{Perception of task-specific self-efficacy}}
We measured participants' perceptions of four sources of self-efficacy using items adapted from \cite{usher2009sources}. The four sources of self-efficacy is based on \cite{bandura1986social}'s social cognitive theory: perception of performance accomplishment, vicarious experience, verbal persuasion, and emotional arousal. Items included ``\textit{After reading the message, I felt more confident in my ability to solve these challenges, based on my past successes}'' (perceived performance accomplishment) and \textit{``After reading the message, I felt more mentally clear and focused when approaching the challenges}'' (perceived emotional arousal). Items were rated on 7-point Likert scales on agreement, and subscale scores were computed as averages. The scale showed excellent internal consistency (Cronbach's $\alpha = .93$).

\paragraph{\textbf{Perception of normative influence}}
Because no validated scales existed for academic task norms, we developed items to measure perception of these normative mechanisms. Items for injunctive and descriptive norms were adapted from \cite{rimal2005how}, reference group identification items were adapted from \cite{cameron2004three}, and social sanction items were developed based on the theoretical framework of \cite{cialdini1990focus} and \cite{lapinski2005explication}, which conceptualize social sanctions as perceived consequences for norm violations. Examples included ``\textit{After reading the message, I believed that solving the challenges with my own skills would be seen positively by the study organizer}'' (injunctive) and ``\textit{After reading the message, I got the impression that most participants solved the challenges on their own}'' (descriptive). Subscale scores were calculated by averaging relevant items on a 7-point Likert scale on agreement. The scale showed excellent internal consistency (Cronbach's $\alpha = .89$).

\paragraph{\textbf{Cognitive discomfort}}
We measured participants' self-reported cognitive discomfort using items adapted from \cite{metzger2020cognitive}. Those items reflect discomfort arising from potential conflicts between messages and task approach. Example items include ``\textit{After reading the message, this challenge made me feel uncomfortable}'' and ``\textit{After reading the message, I disliked the challenge because it challenged my beliefs}'', with some items reverse-coded. Participants rated their agreement on a 7-point Likert scale, and a composite cognitive discomfort score was calculated by averaging responses across all items. The scale showed good internal consistency (Cronbach's $\alpha = .86$).

\paragraph{\textbf{Moral disengagement}}
We assessed participants' self-rated tendency toward moral disengagement (that allows justification of unethical behavior), using items adapted from \cite{shu2011dishonest}. Sample items included ``\textit{Sometimes getting ahead of the curve is more important than adhering to rules}'' and ``\textit{If others engage in bending rules, then the behavior is morally permissible}''. Participants rated their agreement on a 7-point Likert scale, and an overall moral disengagement score was calculated by averaging responses across all items. The scale showed good internal consistency (Cronbach's $\alpha = .87$).

\subsubsection{Qualitative insights into message effects}

To complement the quantitative measures, we included open-ended questions to capture participants' subjective experiences with the intervention messages. Full questions can be found in the Appendix.

\textbf{Message comprehension:} To verify that participants had read and understood the message before the task, they were asked to recall its content (``I remember the message talked about...'') and to rate their comprehension confidence on a 5-point scale ranging from ``Not confident at all'' to ``Completely confident''.

\textbf{Message impact:} To explore emotional and cognitive responses, participants answered ``What emotions or thoughts came up for you while reading the message?'' and reflected on behavioral influence through ``In what specific ways, if any, did the message influence how you approached the word creation challenge?''

\textbf{Message persistence and suggestions:} Participants also reported whether aspects of the message stayed with them during the challenge (``Were there any aspects of the message that stayed with you throughout the challenge?''). They also provided recommendations for improving content (``If you were to create a similar message for future participants, what would you emphasize or include?'').

\subsection{Recruitment and participants}

We recruited 1349 UK-based participants through Prolific.com, a crowdsourcing platform commonly used to collect high quality data from online participants \cite{peer2022data}. Of these, 1282 participants completed the study. However, several participants had incomplete responses. These participants' data was excluded from the final dataset, which thus contained complete data from a total of 1232 participants. Recall that participants in our study were randomly allocated to one of 46 conditions (a control group plus 45 intervention groups), for each of these conditions we have data from approximately 78 participants. Data collection was conducted over a two-week period between June and July 2025. The sample was non-representative, with 51\% female, 47\% male, and 2\% identifying as other/prefer not to say. On average, participants were 29 years old (SD = 4 years) with diverse educational backgrounds. Around 67\% of our participants had a university degree and approximately 36\% of our participants had taken more than 20 online tests in the preceding two years, while 15\% had not taken any.

\subsection{Ethical considerations}

Before the study, participants received a digital information sheet and provided informed consent. Only those who agreed to the tracking of task-related data (e.g., response times, interactions, and anagram-solving patterns) were eligible to participate. The study protocol was reviewed and approved by the university's ethics committee, and all procedures complied with GDPR requirements. Participants were informed about data collection, storage, and their right to withdraw at any time. To maintain experimental validity while measuring cheating behavior, two forms of deception were employed: (1) the true research objectives were withheld during the task, and (2) participants were told performance would determine their bonus payment when in fact all received the maximum bonus. Both deceptions were fully disclosed during debriefing, where participants learned the actual study purpose, the cheating detection methods employed, and the standardized bonus structure. Participants were given the opportunity to withdraw their data after this disclosure. No disciplinary action was taken against participants who cheated, and the debriefing emphasized that the study aimed to understand behavior patterns rather than judge individual conduct.\label{sec:methods}

\section{Results}
We present findings in three main sections that build on one another. The first section tests whether concept-based interventions reduce cheating behavior. The second section examines whether these interventions influence performance and experience across integrity groups (non-cheaters: 0\% cheated; partial-cheaters: some cheated; and full-cheaters: 100\% cheated). The third section then explores the underlying psychological pathways, specifically, how participants' self-reported psychological states or ``mechanisms'' (such as need satisfaction, cognitive discomfort, social norms, and self-efficacy) shape the effects of concept-based messages on cheating, performance, and experience. Understanding these self-reported mechanisms helps explain why interventions work differently across integrity groups.

\subsection{Do concept-based interventions reduce cheating behavior?}

This section addresses RQ1: Do concept-based interventions reduce cheating behavior? We also examined whether the effects on cheating vary across different concepts. To do this, we compared the distribution of integrity groups between the control and intervention conditions using chi-square tests. Additionally, we computed 95\% bootstrapped confidence intervals (1000 iterations) for each concept to quantify the uncertainty around these effects.

\begin{figure}[!ht]
\centering
\includegraphics[width=0.7\textwidth]{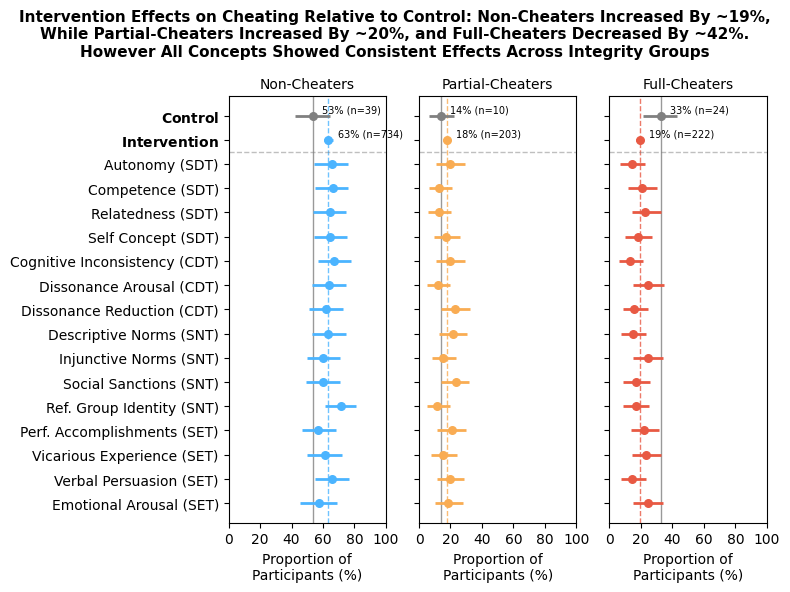}
\caption{Distribution of integrity groups across control and intervention conditions. Forest plots show mean proportions with 95\% bootstrapped confidence intervals (1000 iterations). Points represent means; horizontal lines indicate confidence intervals. Dashed lines indicate the overall intervention mean for reference. SDT includes all concepts related to self-determination, CDT includes all concepts related to cognitive discomfort, SNT includes all concepts related to social norms, and SET includes all concepts related to the self-efficacy framework.}
\label{fig:integrity_distribution}
\end{figure}

As shown in \autoref{fig:integrity_distribution}, intervention conditions had significantly fewer full-cheaters (19\%) compared to control (33\%), a 42\% relative reduction ($\chi^2=7.26, p=.007$). This effect was systematic across all interventions: every individual concept showed a lower proportion of full-cheaters than control, with all 95\% confidence interval upper bounds falling below the control mean. The proportion of non-cheaters was relatively higher ($\sim 19\%$) in intervention conditions (63\%) than control (53\%), while partial-cheaters also increased by 20\% (17\% vs. 14\% in control). However, differences for non-cheaters and partial-cheaters were not statistically significant ($\chi^2<2.50, p>.12$).

\subsubsection{Do the effects on cheating vary by concepts?}

In \autoref{fig:integrity_distribution}, analysis of 95\% bootstrapped confidence intervals (CI) revealed no strong evidence of differential effectiveness across concepts. Substantial overlap of 95\% CI across all individual concepts indicates consistent effects, with concepts' means lying within approximately 11--14 percentage points of the overall intervention mean. All concepts showed directionally similar patterns relative to control, suggesting comparable effects across all theoretical frameworks.

\subsection{Do concept-based interventions affect performance and user experience?}

Prior research suggests that assessment-integrity measures can sometimes reduce performance or negatively affect experiences \cite{alessio2017examining, wuthisatian2020student}. However, these outcomes are often confounded by cheating: when participants cheat, scores no longer reflect true ability, and their experiences may also shift \cite{vanderzanden2018domains}. We therefore analyzed performance and experience separately across the three integrity groups and for all participants combined.

\begin{figure}[!ht]
\centering
\includegraphics[width=0.7\textwidth]{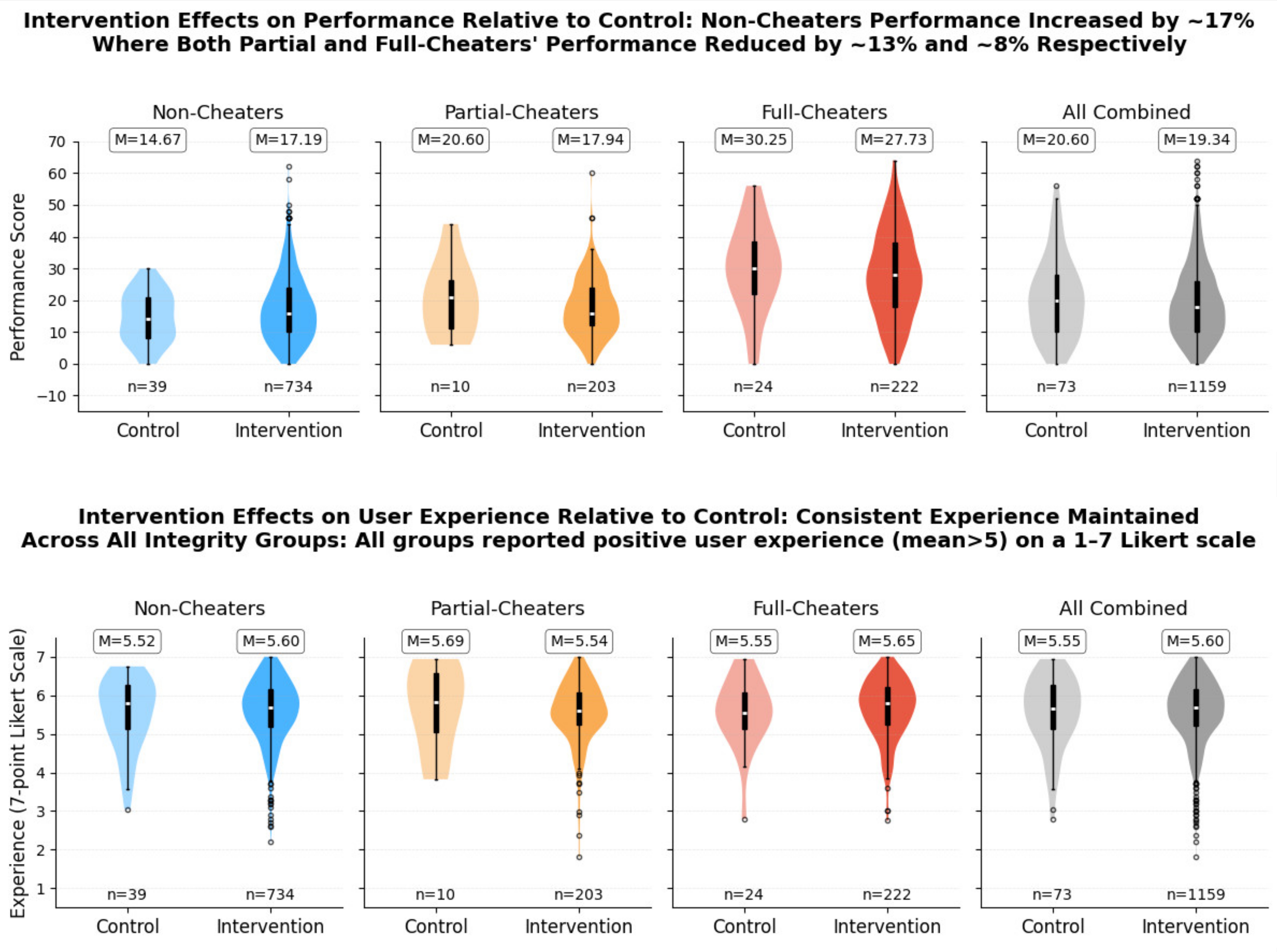}
\caption{Performance and experience distributions by integrity group and condition. Violin plots display data distributions with embedded box plots showing medians (white point) and interquartile ranges. Points with black borders indicate outliers. Sample sizes and means are annotated below and above each distribution respectively. M refers to mean performance of each integrity group under control and intervention.}
\label{fig:performance_experience}
\end{figure}

Beyond examining direct intervention effects, understanding how performance and experience relate is also important. Prior studies have found that students who perform well tend to report higher wellbeing when autonomous motivation prevails \cite{ryan2000self}, but lower wellbeing when moral conflict arises from cheating \cite{harmonjones2019introduction}. Hence, examining the performance-experience relationship can reveal how integrity groups with similar performance may differ in their experiences (e.g., cheaters may feel regret or discouragement after cheating), providing insights toward designing cheater-specific messaging interventions.

We examined RQ2 through three analyses: (1) whether interventions affected performance across integrity groups, (2) whether they affected experience across those groups, and (3) whether the performance-experience relationship differed by integrity group. To compare control and intervention conditions, we used Welch's t-test and the non-parametric Mann-Whitney U test (appropriate for unequal variances and sample sizes; control $n=73$, intervention $n=1159$). To examine performance-experience relationships within each integrity group, we computed Pearson correlation coefficients and compared them using Fisher's Z-tests. \autoref{fig:performance_experience} displays violin plots with embedded box plots showing distributions of performance and user experience.

\subsubsection{Performance effects across integrity groups}

Performance varied substantially by integrity group (see \autoref{fig:performance_experience} for distributions and means). Across all conditions (control and interventions), full-cheaters achieved the highest mean performance scores ($M=28.4$), followed by partial-cheaters ($M=18.5$) and non-cheaters ($M=16.9$), demonstrating that cheating typically inflates performance. The distribution of high scores further illustrates the impact of cheating: about 18\% of full-cheaters scored between 40--50 points, compared to only 1\% of partial-cheaters and none of the non-cheaters. Interestingly, this performance advantage from cheating was similar in both control and intervention conditions.

Comparing control versus intervention conditions within each integrity group revealed modest differences. Non-cheaters showed a 17\% performance increase under intervention conditions (from $M=14.7$ to $M=17.2$), though not statistically significant (Welch's $t=-1.91, p=.063$; Mann-Whitney $U=12,744.5, p=.248$). Partial-cheaters and full-cheaters showed performance decreases of 13\% and 8\% respectively under intervention conditions, but these were also non-significant (Welch's $t<0.85, p>.40$; Mann-Whitney U$, p>.31$).

Research consistently shows that self-rated abilities correlate modestly with objective measures of performance, reflecting people's general awareness of their true competence \cite{zell2014people}. In this study, participants rated their perceived ability to solve anagram puzzles before completing the task. Among non-cheaters, perceived ability correlated positively with performance ($r=0.14, p<.001$), indicating that their self-assessments aligned with actual capability. For partial-cheaters, this correlation was small and negative ($r=-0.12, p=.016$), suggesting a mismatch between perceived and true ability. For full-cheaters, the correlation was nonsignificant ($r=0.07, p=.16$), suggesting that inflated performance scores were disconnected from true competence. These patterns confirm that cheating inflates scores beyond what ability predicts, whereas interventions may help align performance more closely with genuine capability among honest test-takers.

\subsubsection{Experience effects across integrity groups}

Experience ratings remained stable across all integrity groups. Experience mean scores consistently exceeded 5.0 on the 7-point Likert scale used in both conditions, indicating generally positive experiences. Neither Welch's t-tests nor Mann-Whitney U tests revealed significant differences between conditions for any integrity group (Welch's $t: t<0.56, p>.57$; Mann-Whitney U: $p>.45$). Given the absence of significant differences in both performance and user experience between control and intervention conditions, we did not examine variation across individual concepts.

\subsection{RQ3: How do concept-based interventions influence cheating behavior, performance, and experience across integrity groups through participants' self-reported psychological states or mechanisms?}

Previous analyses showed that concept-based messages reduced cheating while maintaining performance and experience; these effects were consistent across all concepts. The next question we want to tackle is how do these interventions actually influence outcomes?

\begin{figure}[!ht]
\centering
\includegraphics[width=0.6\textwidth]{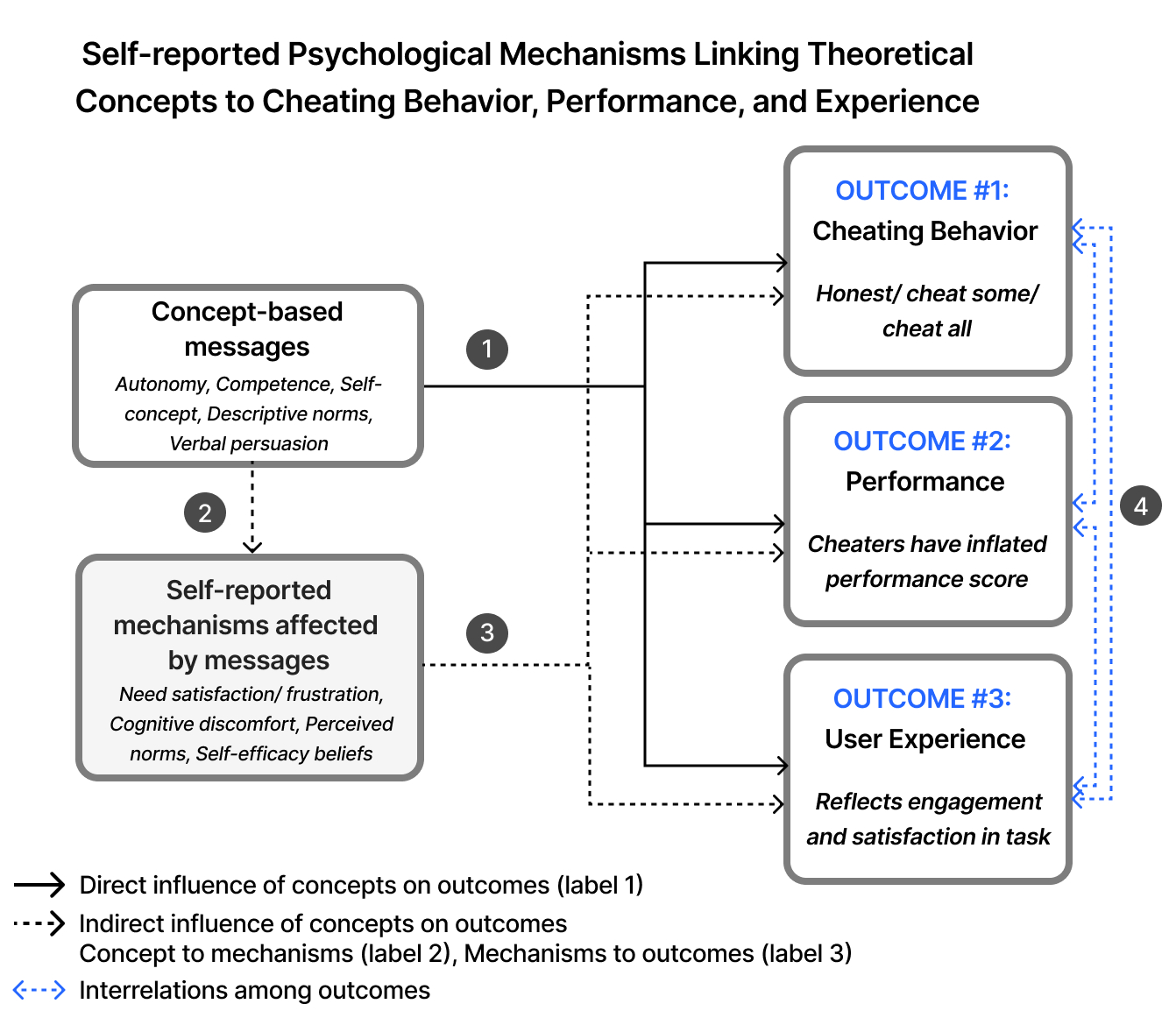}
\caption{Analytical framework showing the different pathways relating concept-based messages to outcomes (cheating behavior, performance, user experience). (1) depicts the direct influence of concepts on outcomes; (2) depicts the potential effects of concepts on self-reported mechanisms; (3) depicts the potential effects of mechanisms on outcomes; and finally, (4) adds an additional interrelations aspect among outcomes.}
\label{fig:analytical_framework}
\end{figure}

\autoref{fig:analytical_framework} illustrates the analytical framework we use for conceptualizing the potential effects of concept-based interventions. Within this framework, concept-based messages can influence outcomes through two pathways: directly from concepts to outcomes (arrows 1), or indirectly through participants' self-reported psychological mechanisms (arrows 2 and 3). The previous sections examined the direct pathway. This section examines the indirect pathway.

Although we designed messages to align with specific theoretical concepts (e.g., autonomy, descriptive norms, verbal persuasion), individuals may process these messages differently \cite{yeager2011social}. A message about autonomy may lead some participants to not only perceive autonomy needs more strongly but it may also affect their sense of social norms or competence needs. It is conceivable that the effectiveness of the messages depends less on the concepts used for their design and more on the internal psychological states or mechanisms (e.g., feeling of need satisfaction) they affect in participants.

To explore this indirect pathway, we examined its two components, corresponding to arrows ``2'' and ``3'' in \autoref{fig:analytical_framework}:

\begin{enumerate}
    \item \textbf{Concept$\rightarrow$Mechanism (arrow 2):} Do the concept-based messages affect the internal psychological mechanisms as intended? For example, does an autonomy message increase perceived autonomy need satisfaction more than other mechanisms? More generally, do messages differ in how they affect those internal mechanisms?
    
    \item \textbf{Mechanism$\rightarrow$Outcome (arrow 3):} Do self-reported mechanisms predict cheating behavior, performance, and experience differently across integrity groups? Determining which specific mechanisms affect those outcomes may further provide insights into how our interventions influenced outcomes.
\end{enumerate}

\subsubsection{Do concept-based messages have the intended effect on self-reported psychological states or mechanisms?}

\begin{figure}[!ht]
\centering
\includegraphics[width=0.8\textwidth]{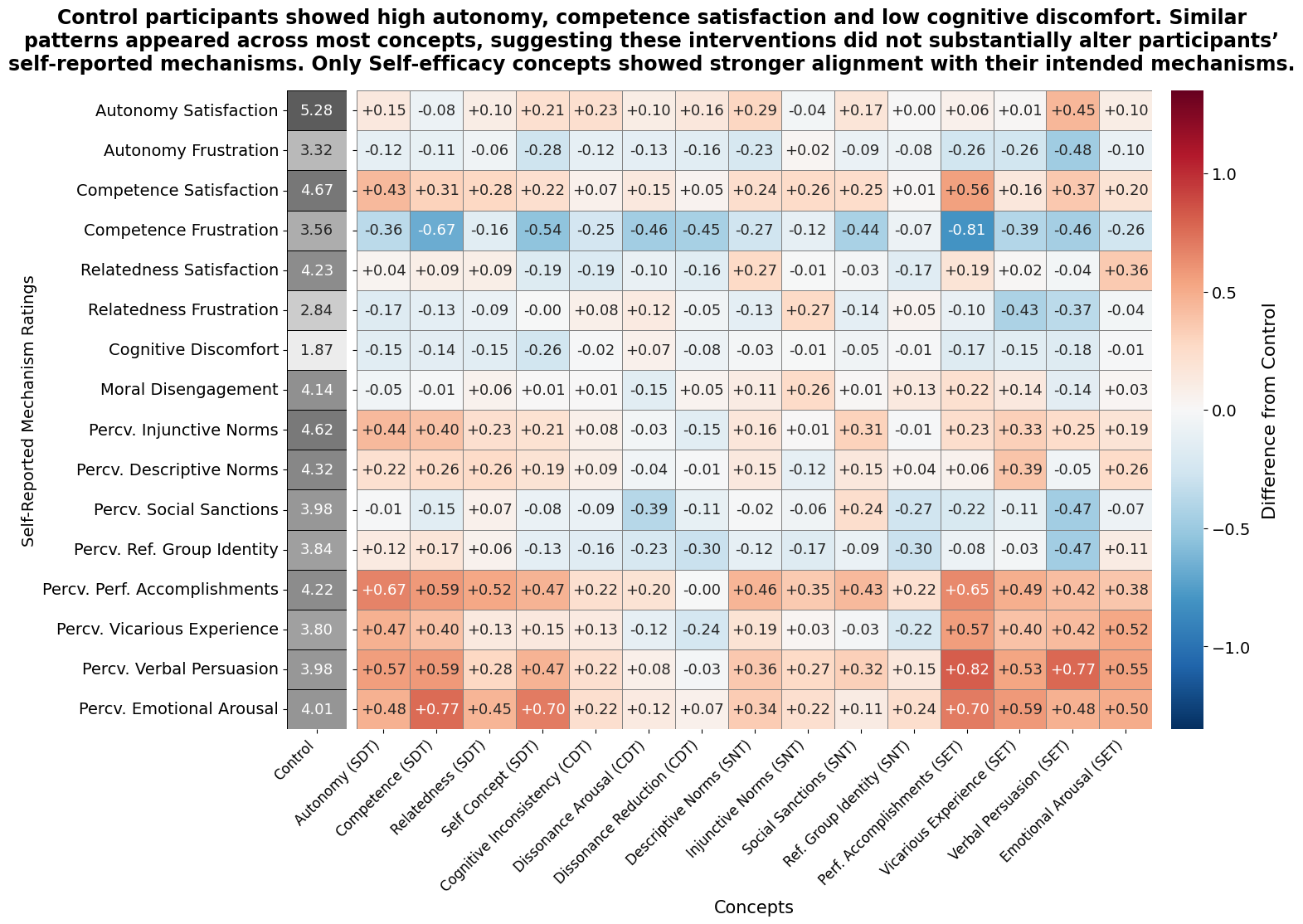}
\caption{Self-reported psychological mechanisms across message concepts. Cells show control means (gray gradients) and deviations for each concept (red/blue), with warmer tones indicating stronger and cooler tones indicating weaker perceived alignment. SDT includes all concepts related to self-determination, CDT includes all concepts related to cognitive discomfort, SNT includes all concepts related to social norms, and SET includes all concepts related to the self-efficacy framework.}
\label{fig:mechanisms_concepts}
\end{figure}

To test the concept-based messages had the intended impact on the targeted internal states or ``mechanisms'' (for example, did autonomy messages increase perceived autonomy need satisfaction more than other mechanisms?), we computed the mean ratings for each mechanism ratings (1-7 Likert scale) for the control group as well as for each concept separately. \autoref{fig:mechanisms_concepts} displays the average mechanisms ratings for the control group on the left-most column. Furthermore, to visualize the effect of each concept on the average rating for each mechanisms, we computed the difference in average rating between a given concept and the control group, such that a value of +0.5 for example, would indicate that a particular concept intervention increased the rating on that mechanism by 0.5 relative to the control group. These concept-specific effects on mechanisms (i.e., differences from control) are displayed in \autoref{fig:mechanisms_concepts}.

Three main patterns can be seen in \autoref{fig:mechanisms_concepts}. First, participants in the control group already reported high autonomy satisfaction (5.28) and competence satisfaction (4.67) but low cognitive discomfort (1.87). This pattern was observed across all concepts. Second, messages had modest effects on self-reported mechanisms: the mean difference from control was 0.07 (SD=0.28), and the highest observed change was +0.82 (approximately 3 standard deviations). Third, the observed effects of the concepts on the mechanisms do not support the intended concept-to-mechanism alignment. Each concept affected multiple mechanisms rather than uniquely affecting its intended mechanism. Self-determination (autonomy, competence, relatedness), cognitive dissonance, and social norms concepts - all showed essentially no effect on their intended mechanisms (all mean $<0.06$). However, self-efficacy concepts (marked SET in \autoref{fig:mechanisms_concepts}) were an exception, producing substantially stronger increases in their four intended mechanisms (range: +0.38 to +0.82, mean=+0.55) compared to other theoretical frameworks.

These results show that concept-based messages may work through broad, multi-mechanism pathways rather than precise concept-to-mechanism targeting. All messages had a psychologically supportive foundation (high autonomy and competence, low discomfort as in control), explaining their consistent effectiveness despite lacking clear theoretical alignment. The lack of one-to-one correspondence suggests that participants' perceptions of the messages through self-reported mechanisms may matter more than the specific theoretical labels assigned to message content when designing effective cheating prevention messages.

\subsubsection{Do self-reported mechanisms predict cheating, performance and experience across integrity groups?}

The previous section established that messages affected multiple internal psychological states or mechanisms at the same time (positively or negatively), unlike the intended concept-to-mechanism alignment. Next we ask whether variations in those self-rated psychological dimensions are associated with differences in the outcomes of interest (arrows 3 in \autoref{fig:analytical_framework}).

\paragraph{\textbf{Do mechanisms predict cheating behavior?}}
To test whether self-reported psychological states or mechanisms can predict cheating behavior, we ran separate logistic regression models for each integrity group (i.e., predicting non-cheaters vs others, partial-cheaters vs others, and full-cheaters vs others) using as input participants ratings across the 16 mechanisms.

First, to evaluate if mechanisms can predict cheating behavior, we compared each full model (including all mechanisms) against a null model (including no mechanisms; intercept only) using likelihood ratio tests. All three models significantly improved prediction compared to the null, indicating that self-reported mechanisms as a set contain meaningful information about cheating behavior (non-cheaters vs others: $\chi^2(16)=74.82, p<.001, \text{pseudo-}R^2=.024$; partial-cheaters vs others: $\chi^2(16)=27.91, p=.032, \text{pseudo-}R^2=.013$; and full-cheaters vs others: $\chi^2(16)=112.17, p<.001, \text{pseudo-}R^2=.047$. Pseudo-$R^2$ values being between .013 and .047 indicate small but reliable effects: mechanisms are predictive of cheating behavior, but they explain only a modest proportion of behavioral differences.

Next, to assess which mechanisms matter most, we used permutation importance, which estimates each mechanism's contribution to model accuracy. Because it captures predictive contribution rather than direction, permutation importance indicates which mechanisms most influence classification, regardless of whether they increase or decrease the likelihood of cheating behavior. This analysis shows that all importance values were very small (importance value $<0.004$), meaning that no single mechanism strongly distinguished the groups. Instead, predictive performance relied on subtle, overlapping effects of many interrelated mechanisms. This suggests that cheating behavior is shaped by patterns across multiple processes, rather than a few dominant predictors.

\begin{figure}[!ht]
\centering
\includegraphics[width=1\textwidth]{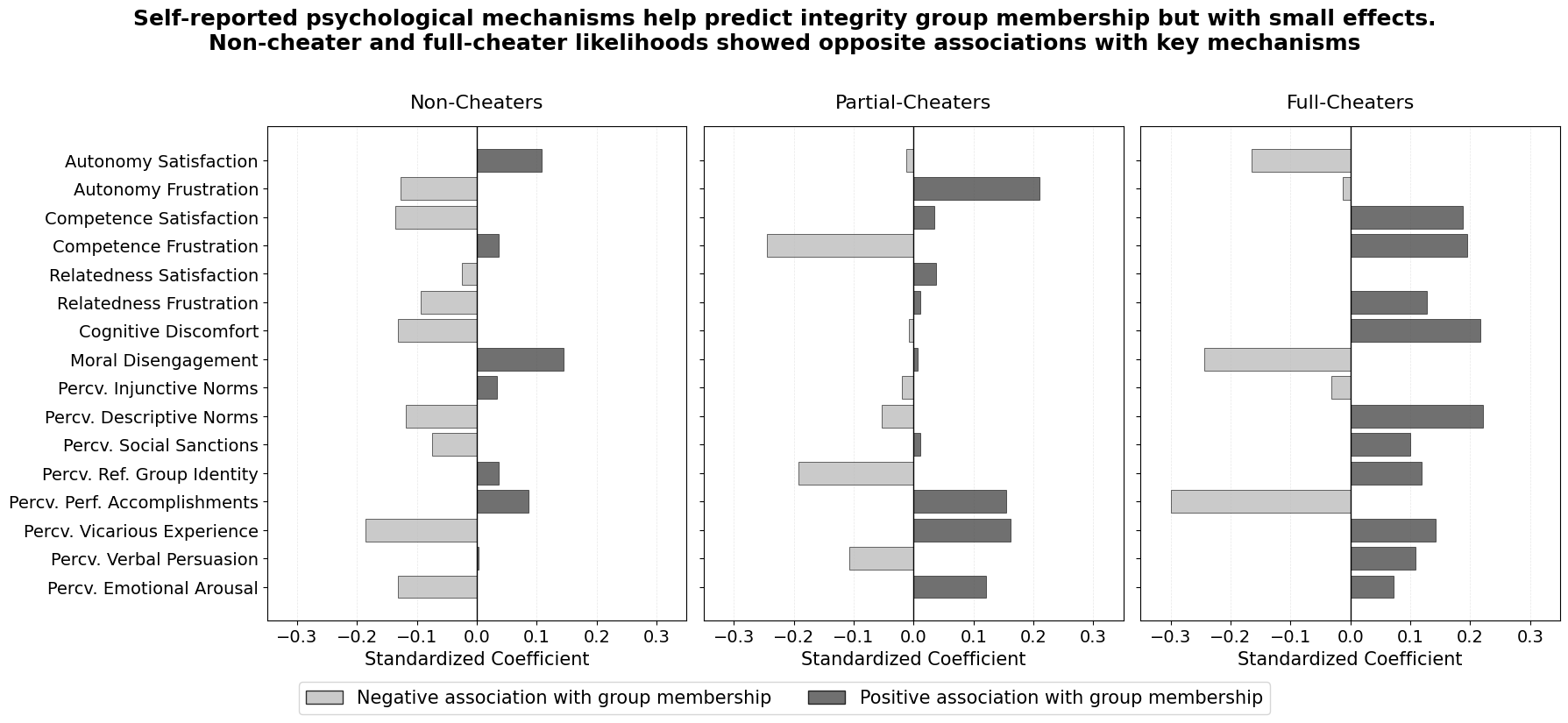}
\caption{Self-reported Psychological Mechanisms Associated with Integrity Group Membership. Separate binary logistic regression models compared non-cheaters vs others (left), partial-cheaters vs others (middle), and full-cheaters vs others (right). Dark gray and light gray bars indicate positive and negative associations with group membership respectively. Coefficients are standardized but should be interpreted cautiously due to multicollinearity among mechanisms ($\sim 50\%$ of mechanism pairs: $|r|>0.3$). All models significantly outperformed null models ($p\leq.032$).}
\label{fig:mechanisms_cheating}
\end{figure}

Finally, to assess how self-reported mechanisms differ across groups, we investigated standardized regression coefficients to understand directionality, that is, whether higher self-rated psychological states or mechanisms were associated with a greater or lower likelihood of belonging to each integrity group. These results are shown in \autoref{fig:mechanisms_cheating}. Higher perceived injunctive norms, emotional arousal, and competence frustration were associated with greater likelihood of being a non-cheater, while higher cognitive discomfort and autonomy satisfaction were associated with lower likelihood of being a non-cheater. Full-cheater group membership showed roughly opposite patterns; with higher cognitive discomfort and autonomy frustration, and lower perceived norms, emotional arousal, and competence frustration were associated with higher likelihood of being a full-cheater. Interestingly, lower perceived performance accomplishments were associated with greater likelihood of being a partial-cheater. If mechanisms were uncorrelated, these coefficients would represent independent effects. However, because mechanisms are correlated ($\sim 50\%$ of pairs have $|r|>0.3$), the directions are informative about relative tendencies within each psychological profile rather than isolated causal predictors.

Taken together, Pseudo-$R^2$ values show that self-reported mechanisms as a group contribute modestly to predicting cheating. Permutation importance shows that no individual mechanism dominates prediction. Regression coefficients show the directional psychological patterns that distinguish the groups. Overall, the small effect sizes highlight that cheating behavior likely depends on broader situational and individual factors beyond the mechanisms measured here.

\begin{figure}[!ht]
\centering
\includegraphics[width=0.8\textwidth]{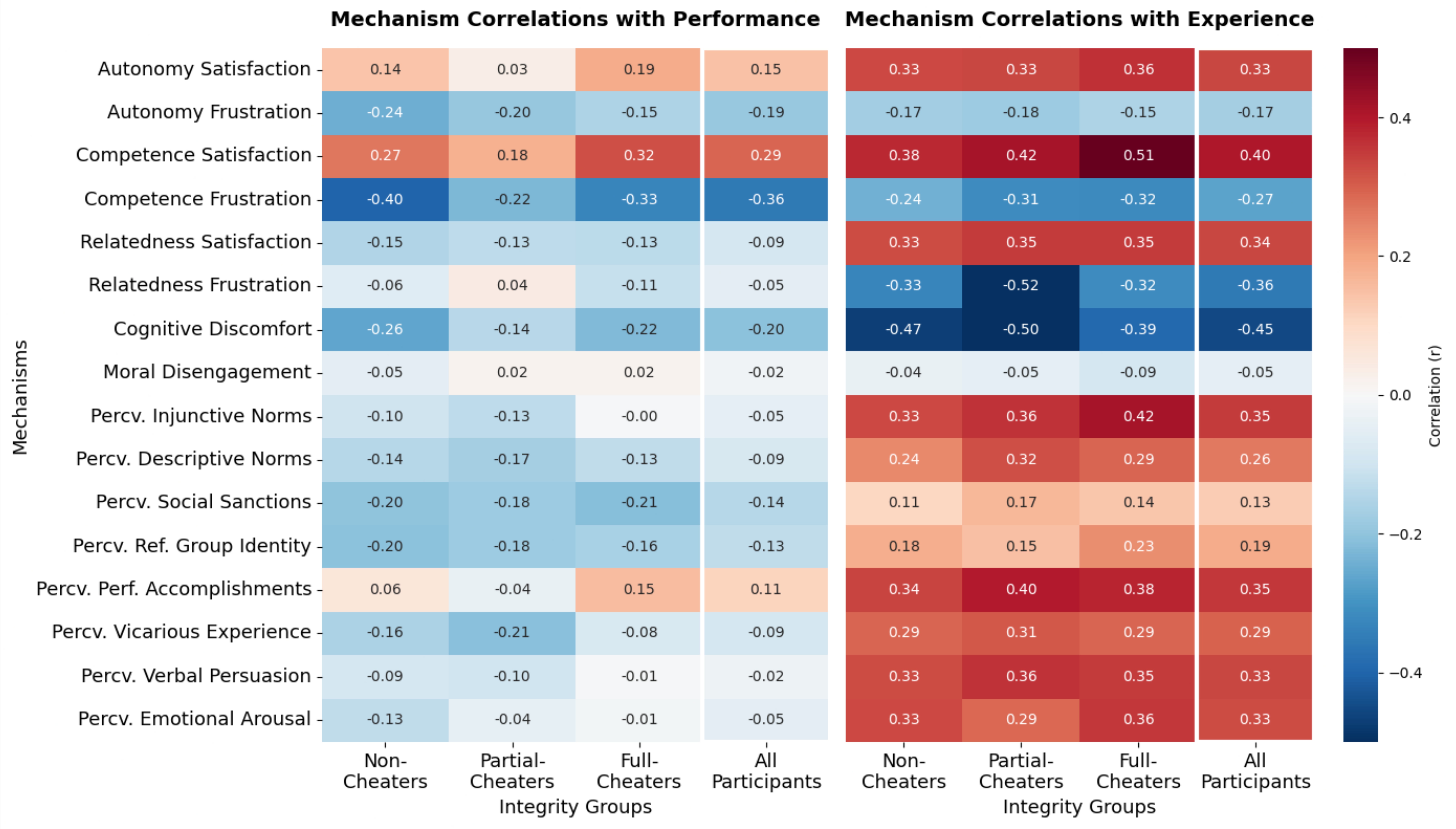}
\caption{Correlations between self-reported psychological mechanisms and performance (left) and experience (right) across integrity groups. Heatmaps display Pearson correlation coefficients (red=positive, blue=negative). Correlations with experience were larger in magnitude than with performance and mostly positive. Competence satisfaction positively correlated with both performance and experience across all groups. Social norms (e.g. perceived social sanctions, perceived group identity) showed contrasting patterns: negatively associated with performance but positively associated with experience, while cognitive discomfort and relatedness frustration strongly undermined experience.}
\label{fig:mechanisms_outcomes}
\end{figure}

\paragraph{\textbf{Mechanism-performance correlations}}
To assess the relationship between variations in self-reported mechanisms and performance we computed Pearson correlations between mechanisms and performance. We computed these correlations separately for each integrity group and for all participants combined because cheating behavior affects performance. These correlations are shown in \autoref{fig:mechanisms_outcomes} (left panel). Several points are worth noting here. First, the correlations between performance and mechanisms are quite similar across integrity groups. Second, most correlations are rather small and negative. Third, some correlations are quite large and may thus have practical and theoretical implications. For instance, competence satisfaction correlates positively with performance across all groups (non-cheaters: $r=0.27$; partial-cheaters: $r=0.18$; full-cheaters: $r=0.32$, all $p<.001$), while competence frustration correlates negatively with performance (for non-cheaters, $r=-0.40, p<.001$). Social norms mechanisms (i.e., perceived social sanctions, perceived reference group identity) negatively correlated with performance for both non-cheaters and cheaters ($r$ ranges between $-0.21$ and $-0.16, p<.001$).

\paragraph{\textbf{Mechanism-experience correlations}}
To assess the relationship between self-reported mechanism and experience we used the same approach and computed the same type of correlations, which are displayed on the right side of \autoref{fig:mechanisms_outcomes}. Several points are worth noting here.

As with performance, the effects seem to be very similar across integrity groups. However, the correlations are of larger magnitude than they were for performance and they are mostly positive. Some mechanisms have a strong positive correlation with experience. For example, competence satisfaction correlates strongly with experience across all groups (more than performance), strengthening across the integrity spectrum (non-cheaters: $r=0.38$; partial-cheaters: $r=0.42$; full-cheaters: $r=0.51$, all $p<.001$). Social norms and self-efficacy related mechanisms positively correlate with experience (most $r>0.30, p<.001$). Other mechanisms have a strong negative correlation with experience. Cognitive discomfort and relatedness frustration negatively correlate with experience, strongest for partial-cheaters ($r\approx -0.50, p<.001$).

Taken together, the indirect pathway (arrows ``3'' in \autoref{fig:analytical_framework}) is important to understand how message interventions affect outcomes of interest. While self-reported mechanisms as a group contain meaningful information about cheating behavior across integrity groups (all $p<.04$), their contribution is modest (pseudo-$R^2$ ranging from .013 to .047), with no single mechanism dominating prediction. The mechanisms reveal distinct psychological profiles characterizing each integrity group rather than strong causal predictors, suggesting that cheating behavior depends on broader situational and individual factors beyond the self-reported mechanisms measured here. Mechanisms seem to relate mostly to experience and somewhat to performance. We also observed that competence satisfaction has positive correlations with both performance and experience across integrity groups.

\subsection{How do performance and experience relate across integrity groups?}

Past research has shown that increased performance can enhance wellbeing and that increasing wellbeing can enhance performance \cite{vanderzanden2018domains}, suggesting reciprocal positive relationships. Hence, in this section, we explore how performance and experience correlate across integrity groups.

We calculated Pearson correlations between performance and experience within each integrity group using all participants (control and intervention combined). Performance-experience correlations differed substantially across integrity groups: non-cheaters showed weak positive correlation ($r=0.12, p<.001$), full-cheaters showed moderate positive correlation ($r=0.22, p<.001$), and partial-cheaters showed no significant correlation ($r=0.01, p=.89$). Fisher's Z-tests confirmed these differences were statistically significant (all $p<.05$).

These findings show that performance and experience relate differently across integrity groups suggesting distinct patterns across integrity groups. Non-cheaters show a weak coupling between experience and performance. One potential explanation for this result is that non-cheaters may derive their satisfaction from multiple sources beyond performance, such as maintaining integrity and autonomous motivation. Full-cheaters show a stronger coupling between performance and experience because better performance achieved through cheating may directly enhance their sense of success. Partial-cheaters show no correlation, possibly reflecting tension where performance gains could have failed to produce experiential benefits due to conflicting feelings about their approach.

\subsection{Understanding message comprehension and perceived impact: what did participants actually experience?}

To gain a deeper understanding of participants' experience and perception of the messages, we asked them a series of open-ended questions about what they remembered, their emotional responses, and their perceived impact on task behavior (see the Appendix for full questions). Note that participants were not asked to identify specific internal psychological states or mechanisms in the open-ended questions.

Using inductive coding in MAXQDA (ver. 2024), the first author systematically categorized responses into broader themes \cite{creswell2016qualitative}. Detailed coding results are in the Appendix. Content analysis shows that the message comprehension was moderate and similar across integrity groups. Approximately 60--66\% of participants exhibited at least partial understanding of the content of the message, while 21--26\% expressed confusion about the purpose of the message. Emotional responses were balanced, with roughly 27\% of participants reporting positive feelings and 10\% reporting negative feelings across all integrity groups.

Most participants reported that the message they were shown did not influence how they approached the task: 56\% of non-cheaters, 43\% of partial-cheaters, and 39\% of full-cheaters said the messages did not change how they solved anagrams. Roughly 20--30\% across integrity groups reported explicitly thinking about messages during the task.

Next, we observed how integrity groups recalled and engaged with messages varied by concept type. Full-cheaters most accurately recalled control and capability messages (93\% for self-concept, 85\% for social-sanction) and reported finding autonomy messages most motivating (45\%). Partial-cheaters most accurately recalled moral tension messages (89\% for dissonance-arousal) and reported finding verbal persuasion most motivating (53\%). Non-cheaters most accurately recalled encouragement and peer-comparison messages, with verbal persuasion showing highest recall accuracy (84\%) and vicarious experience being rated as most motivating (42\%).

When asked for suggestions to improve messages, around 50\% across all groups requested more content that would motivate them to perform well, while only 18--21\% requested more task-specific information.

Overall, participants generally understood the messages and reported mild emotional reactions, with engagement and recall varying across integrity groups. Participants preferred content that enhanced motivation, and for some, the messages influenced reflection and task-related motivation. Such qualitative insights complement the quantitative findings by highlighting how different messages are attended to and perceived across individuals.\label{sec:result}

\section{Discussion}
Valid remote, unproctored assessment depends on both integrity and fairness. Students or test-takers should not gain unfair advantages through dishonesty, but they also deserve conditions that allow them to exhibit what they truly know. These two goals, honesty and fairness, are not opposites; they often work together to make assessments meaningful \cite{mukherjee2023effects, chin2017influence}. However, many current message-based interventions to prevent cheating in assessments have important weaknesses. They often lack a solid theoretical foundation, resulting in inconsistent effect sizes; focus primarily on preventing cheating; rely on warnings or deterrent messages that can increase stress and negatively affect performance rather than fostering honest and motivated engagement; and maintain a simplistic, black-or-white view of cheating.

This study aimed to systematically address these gaps by leveraging four psychological theories to design 15 concept-based interventions framed in motivational, rather than deterrence-oriented, language, and examined their effects across an integrity spectrum on cheating, performance, and experience. Below, we discuss: the effects of theory-driven design in Section~\ref{sec:theory-design}; the role of motivational framing and the preservation of both performance and user experience in Section~\ref{sec:motivational-framing}; the distinct psychological behavior across the integrity spectrum in Section~\ref{sec:integrity-spectrum}; and a systematic analysis of mechanisms, showing how self-reported underlying psychological processes can predict outcomes differently across integrity groups in Section~\ref{sec:mechanisms-pathways}. Finally, we situate these findings within a broader context and discuss their practical and theoretical implications in Section~\ref{sec:implications}.

\subsection{Theory-grounded systematic design produces comparable effects}
\label{sec:theory-design}

Previous research on integrity interventions has shown mixed and inconsistent results. For example, honor codes or warning messages sometimes reduce cheating, but the effects vary widely (from 20\% to 80\%; \cite{bing2012experimental, mccabe1999academic, corrigangibbs2015deterring, pleasants2022cheating}). One potential reason for these inconsistent results is that most messages are developed in an ad hoc way and lack a clear theoretical foundation. This makes it difficult to understand which interventions would work and to make systematic progress in this field.

In this study, we adopted a theory-driven approach to design messages grounded in 15 psychological concepts. Despite their grounding in different psychological theories, all messages led to similar reductions in full cheating (from 33\% in the control condition down to 19\% in the intervention condition, a 42\% relative reduction). This shows that when messages are designed following a systematic approach, their effects become more predictable and consistent. However, the absence of differences in effectiveness across concepts (i.e., autonomy, competence, cognitive dissonance, social norms, and self-efficacy) is somewhat surprising. Each of these concepts has been theorized to promote honesty through a distinct pathway, yet all contributed equally to fostering a supportive atmosphere for test-takers in our study.

Importantly, our results show that our interventions mainly reduced full cheating but not partial cheating. This suggests that while messages can discourage complete dishonesty, smaller lapses may require different strategies. Overall, a theory-based approach allows institutions to select message types that best fit their values while achieving reliable outcomes, replacing trial-and-error approaches with systematic design.

\subsection{Motivational framing reduces cheating while preserving performance and experience}
\label{sec:motivational-framing}

Traditional interventions face two main challenges. First, they rely on deterrence, warnings, surveillance cues, and moral threats, which may reduce cheating but often create stress that undermines honest test-takers' performance \cite{chin2017influence, wuthisatian2020student}. Second, they focus narrowly on cheating reduction, overlooking potential negative effects on performance or the overall user experience.

We intentionally reframed message content from deterrence to motivational support, drawing on studies that link autonomy support and the fulfillment of psychological needs to reduced academic dishonesty \cite{pulfrey2019under, kanatmaymon2015role, bureau2014parental}. Messages avoided explicit references to cheating or honesty, emphasizing instead capability, autonomy, and belongingness to a group of honest peers. This approach reflects a core motivational principle: integrity is more likely to emerge naturally when individuals feel competent, autonomous, and connected \cite{ryan2000self}.

Our findings support the feasibility of this approach. Systematic cheating behavior decreased substantially, while performance and experience were unaffected. Non-cheaters appeared to benefit particularly from the interventions, showing signs of enhanced performance. Moreover, the relationship between self-reported ability to solve anagrams before the task and actual performance was stronger for non-cheaters, suggesting that motivational messages can help honest individuals perform in a way that more accurately reflects their true capabilities.

These results challenge the assumption that integrity interventions must rely on deterrence. Unlike prior approaches emphasizing surveillance or moral threat \cite{corrigangibbs2015deterring, mukherjee2023effects, pleasants2022cheating}, motivational framing sustains honesty through intrinsic rather than extrinsic motivation. Hence, by fostering intrinsic motivation, positive self-concept, supportive norms, and self-efficacy, interventions can reduce cheating while preserving conditions that allow honest test-takers to thrive.

\subsection{Integrity spectrum reveals distinct psychological profiles}
\label{sec:integrity-spectrum}

Most prior research treats cheating as a binary behavior: students either cheat or do not \cite{corrigangibbs2015deterring, mukherjee2023effects}, potentially masking meaningful variation in integrity behaviors and psychological experiences. Our integrity-spectrum classification (non-cheaters, partial-cheaters, full-cheaters) reveals distinct profiles in self-reported perceptions and experiences.

Non-cheaters reported higher injunctive norms and emotional arousal but lower cognitive discomfort, suggesting a psychological profile in which honest behavior can potentially align with both personal and social values. Partial-cheaters reported elevated verbal persuasion and competence-related mechanisms (both satisfaction and frustration), alongside reduced relatedness satisfaction, reflecting psychological tension during task engagement. Full-cheaters were characterized by elevated cognitive discomfort and stronger perceptions that peers cheat (descriptive norms), coupled with lower injunctive norms, consistent with contexts in which dishonesty feels normalized.

Three methodological considerations are worth noting. First, our between-subjects design means that group membership reflects different individuals rather than changes within participants; this prevents us from understanding which specific types of people were affected by the messages, and we can only make claims at the group level. Second, self-reported psychological states (``mechanisms'') were measured post-task, meaning that the experience of the task may have affected those states in addition to the messages. Third, in our analysis, the correlations between concepts, mechanisms, and outcomes assume a particular causal structure (as shown in \autoref{fig:analytical_framework}) and do not allow us to infer causality. Despite these limitations, the patterns of results clearly indicate that self-rated psychological states systematically differ across integrity groups.

These findings extend \cite{bandura2016moral}'s moral disengagement theory by demonstrating that cheating involves multifaceted psychological experiences rather than single-factor deficits. Unlike prior research emphasizing individual moral failings \cite{mccabe1999academic, pascualezama2020cheaters}, the integrity spectrum shows that partial-cheaters constitute a distinct psychological group from full-cheaters. Practically, this suggests that broad-based psychological support, such as preserving autonomy, fostering competence, and building connection, might be more effective than targeting narrowly defined ``at-risk'' profiles, as all participants benefited from motivational messages regardless of their cheating behavior.

\subsection{Mechanisms as explanatory pathways: do self-rated internal psychological states better predict outcomes?}
\label{sec:mechanisms-pathways}

Beyond addressing the four shortcomings listed above, we also conducted a systematic investigation of how participants' self-rated psychological states (``mechanisms'') relate to intervention outcomes across integrity groups, suggesting explanatory pathways largely unexplored in prior research.

Theoretical considerations would suggest that interventions should target specific mechanisms (e.g., autonomy messages $\rightarrow$ autonomy satisfaction $\rightarrow$ honest behavior); yet our findings highlight that, in reality, things are more complicated. Messages often influenced multiple mechanisms simultaneously, challenging assumptions about precise targeting. This result indicates that test-takers appear to process messages through multiple psychological lenses at once, suggesting that interventions operate holistically rather than specifically.

Despite weak concept-to-mechanism correspondence, self-rated mechanisms predicted outcomes differently across integrity profiles. For example, higher feelings of competence satisfaction tended to support increased performance and more positive task experiences, whereas stronger feelings of frustration with competence and relatedness predicted poorer outcomes, particularly for partial cheaters. Stronger self-reported social norms mechanisms (perceived social sanctions and reference group identification) were sometimes linked with reduced performance (particularly for non-cheaters) but positively associated with user experience, illustrating nuanced trade-offs that remain invisible when considering a single outcome.

Three key interpretations emerge. First, post-task self-reports may not reflect real-time decision-making; they could capture post-hoc reflections rather than mechanisms that actually influenced behavior. Second, mechanisms seem to operate collectively, yet all messages effectively supported autonomy, competence, and low cognitive discomfort, creating a broadly supportive psychological environment. This aligns with \cite{walton2018wise}'s `wise intervention' framework, suggesting that messages helped participants view the assessment as autonomy- and competence-supportive, promoting honest engagement regardless of the emphasized concept. Third, the uniform effectiveness of messages despite weak concept-mechanism links implies that fostering broad psychological support may be more robust than targeting specific pathways.

This study also has practical implications. It suggests that any theoretically grounded message that preserves psychological support may effectively reduce cheating. However, it is important to note that our measurements and models explained only a small portion of the variance, and that the causal or temporal relationships assumed in this study remain uncertain. It is therefore possible that important factors were not considered, which might shed a different light on understanding message interventions. Furthermore, future research using experience-sampling to capture real-time internal states could clarify the dynamics of psychological states during the test-taking experience and their impact on ethical decision-making.

\subsection{Implications, limitations and future directions}
\label{sec:implications}

Our findings suggest that academic integrity is shaped not solely by individual morality, but also by how the assessment context is framed through messaging. Integrity thrives in contexts that promote fairness, trust, and autonomy, rather than those dominated by surveillance, fear, or punishment \cite{mazar2006dishonesty, trevino2014ethical}. Supportive messages consistently reduced cheating, irrespective of the specific theories on which they were grounded, suggesting that tone and framing may matter more than content. Importantly, our study leveraged AI-assisted message design to generate theoretically informed interventions, demonstrating how technology can support precise, scalable, and tailored integrity messaging. In practice, this implies that interventions should use motivational language, preserve psychological support, and integrate into broader institutional culture rather than relying on one-off warnings. By fostering both honesty and well-being, such interventions can enhance students' performance while maintaining fairness in assessments.

Our study has several limitations. First, the between-subjects, low-stakes online design prevents tracking individual integrity trajectories. For example, although the intervention reduced the overall percentage of full-cheaters, we cannot determine which participants changed their intentions or how they would have been classified without the message. Second, we focused on comparing message types, but within-subject designs could provide richer insights into how interventions shape behavior over time. Third, we collected self-reported psychological states only at the end of the task to avoid disrupting natural engagement, and because participants needed to experience the full task to reflect on the message's influence. However, post-task responses may be biased by performance or rationalization rather than real-time feelings. Fourth, classifying participants into three integrity groups is an improvement over a binary cheater/non-cheater approach, but it likely misses subtler differences, particularly among partial cheaters. Finally, the controlled, supportive, game-like anagram task differs from most high-stakes test-taking contexts, limiting the generalizability of our findings.

Future research should address these limitations through longitudinal or within-subjects designs that track integrity over time, experience-sampling and multimodal measures capturing real-time psychological states, and fine-grained behavioral analyses to detect nuanced integrity profiles. Investigating additional mechanisms, their interactions, and individual differences, such as personality traits, prior cheating history, or cultural values, can also provide a deeper understanding of why interventions succeed or fail. It goes without saying that cross-cultural studies are particularly important, as psychological needs, social norms, and self-efficacy beliefs vary across contexts, potentially moderating intervention effectiveness. The long-term effects of motivational messages also remain unknown; repeated exposure may be necessary to foster lasting integrity habits.

In sum, our study illustrates that effective interventions do more than reduce dishonesty; they also promote psychological safety, fairness, and motivation, which together strengthen both integrity and performance. By moving beyond deterrence approaches and binary definitions of cheating, and instead fostering context-sensitive, theoretically informed strategies, this work provides a foundation for honest, fair, and psychologically supportive educational environments.\label{sec:discussion}

\section{Conclusion}
This study advances our understanding of academic integrity interventions in remote, unproctored assessments. We show that brief, theory-aligned motivational messages can substantially reduce cheating without impairing performance or user experience, demonstrating that integrity and fairness can be enhanced without relying on surveillance or deterrence.

By addressing four key shortcomings, ad hoc message design, a narrow view of cheating behavior, deterrence-based framing, and binary conceptualization, this study shows that theory-grounded motivational messages can reduce full cheating by 42\% without impairing performance or user experience. Three conceptual advances emerge. First, by adopting a systematic design approach, we created messages grounded in 15 psychological concepts drawn from four major psychological theories; all messages were equally effective at reducing cheating. Second, the study highlights motivational framing over deterrence: messages emphasizing capability, autonomy, and belonging reduced cheating without adding stress or impairing performance. Third, it argues for treating integrity as a multifaceted, rather than binary, construct: non-, partial-, and full-cheaters exhibited distinct psychological profiles, and because the proportions of these groups were influenced by messages, cheating appears to depend not only on moral failure but also on context.

This study also examines how self-reported psychological states (``mechanisms'') after exposure to concept-based messages (``concepts'') can predict outcomes across integrity groups. While concept-to-mechanism links were weak, mechanisms were predictive of cheating, performance, and experience, particularly strongly for experience. Higher levels of self-reported competence satisfaction were associated with both improved performance and user experience, while higher levels of self-reported cognitive discomfort were associated with worse user experience. Higher levels of self-reported social norms (e.g., perceived reference group identification) fostered belonging, despite minor performance costs for honest-takers. These results provide insights into why interventions may succeed and how different profiles of participants may respond differently to the same messages.

From a practical perspective, these findings offer institutions realistic alternatives to deterrence-based approaches, especially as AI-based methods for detecting cheating become less reliable. Motivational messages that preserve or promote a psychologically supportive environment may reduce cheating without harming honest test-takers. From a theoretical perspective, we argue for shifting the focus from individual morality to contextual affordances, as trust-building and autonomy support, rather than enforcement, form the foundation of valid assessment. Integrity and fairness are not opposites but complementary objectives that can be achieved when interventions create environments where honest behavior feels natural, supported, and valued.\label{sec:conclusion}

\bibliographystyle{ACM-Reference-Format}
\bibliography{8_references}

\break

\appendix
\section*{Appendix}

\appendix

\section{List of messages created as interventions}
{\footnotesize 
\begin{longtable}{|p{3cm}|p{1.5cm}|p{0.8cm}|p{8cm}|}
\caption{Final motivational messages selected in the expert workshop for each psychological concept across theories. Three highest-rated messages per concept based on concept alignment and motivational potential}
\label{tbl:messages_appendix} \\
\hline
\textbf{Theoretical Framework} & \textbf{Concept} & \textbf{Msg \#} & \textbf{Created Message} \\
\hline
\endfirsthead
\hline
\textbf{Theoretical Framework} & \textbf{Concept} & \textbf{Msg \#} & \textbf{Created Message} \\
\hline
\endhead
\hline
\endfoot

Self-Determination & Autonomy & 1 & Are you aware that it's up to you how you tackle these challenges? You have the choice in how to approach them, how much effort to put in, and how much you value the results. The way you behave and tackle these challenges, therefore, reflects your personal choices. \\

Self-Determination & Autonomy & 2 & How do you feel when you're doing something that's truly your own? Even if it's not perfect, it still reflects your values and who you are. That sense of ownership keeps you true to yourself---because you're the one who decides how it happens. \\

Self-Determination & Autonomy & 3 & You're in the midst of a tough project, and it's easy to feel stuck. But here's the thing: when you get to decide how you tackle it, you're more likely to own it. It's like you're the captain of your own ship, and that feeling of being in charge makes all the difference. \\

Self-Determination & Competence & 1 & You know that feeling when you're working on something and it finally clicks? It's like you've unlocked a new level of understanding. Can you tap into that feeling and trust that you have what it takes to figure things out, even when it's tough? \\

Self-Determination & Competence & 2 & Can you think of a time when you felt competent doing a challenging task? Challenging tasks can be frustrating at first, but it's amazing how hard work can help you grow and become more capable. \\

Self-Determination & Competence & 3 & Imagine that you are stuck on one challenge, and it is really testing you. What do you do when you hit a roadblock - do you give up, or do you keep trying, learning from what did not work last time? When you persist and finally figure it out, you will realize that it was the process of trying, failing, and trying again that made you better at it. \\

Self-Determination & Relatedness & 1 & Can you think about all the people that are impacted by how you behave? When you feel like your work is genuinely connected to others, it's amazing how much more meaningful it becomes. \\

Self-Determination & Relatedness & 2 & What's the best part of having people who genuinely care about you? It's like having a safety net---you feel more confident to take risks and be your best self. Knowing they're there to support you makes it feel like you can handle anything that comes your way. \\

Self-Determination & Relatedness & 3 & Imagine being how it is like when you are surrounded by people you deeply care about? You know, the ones who make you feel like you are home? When you are with them, you're more likely to open up and share your ideas, and that is when the magic happens and you are building a true connection with others. \\

Cognitive Dissonance & Self-Concept & 1 & So, you're working on something that's really pushing you. Can you think about what makes you, you - like, what sets you apart from others? When you make choices that honor those unique qualities, you feel like you're on the right path, like you're being true to yourself. \\

Cognitive Dissonance & Self-Concept & 2 & Are you noticing how you approach problems in a way that feels uniquely yours? It's like you're drawing from a personal compass that guides your decisions. Trust that inner guide and let it steer you towards solutions that feel authentic and true to who you're becoming. \\

Cognitive Dissonance & Self-Concept & 3 & What does it mean when your work feels authentic? It's like putting your stamp on it---saying, 'This is me, this is mine.' When you leave a piece of yourself in what you do, your work truly becomes meaningful and in return, it shapes and strengthens who you are. \\

Cognitive Dissonance & Cognitive Inconsistency & 1 & What's your gut telling you about how you're approaching these problems? Do you sense any contradictions in the ways you think or feel about your approach? It's worth reflecting on what's important to you and to act accordingly. \\

Cognitive Dissonance & Cognitive Inconsistency & 2 & Don't you ever find yourself caught between two mindsets when working on something? One part of you says, 'I want to do this right,' while the other says, 'Just get it done'. These opposing approaches can coexist and it's interesting to think about how they can shape the way you work. \\

Cognitive Dissonance & Cognitive Inconsistency & 3 & You're really invested in this project. What's going on when you switch between being meticulous and just getting things done? It's like you're balancing two different mindsets, and that's pretty interesting to notice. \\

Cognitive Dissonance & Dissonance Arousal & 1 & What happens when you're working on something and you don't quite give it your all? You're gonna feel it, right - that uneasy feeling that you're not being true to yourself? It's like your actions and your values are at odds, and that tension is gonna stick with you until you find a way to make things right. \\

Cognitive Dissonance & Dissonance Arousal & 2 & What is the first thing that comes to mind when you are proud of what you have accomplished? You feel good, right? But when you did something that does not quite live up to your own standards -- doesn't that feeling of pride turn into something else, like a nagging voice that won't let you move on? \\

Cognitive Dissonance & Dissonance Arousal & 3 & What do you think happens when you look back on a project and realize it doesn't quite reflect who you are? Don't you feel like something's off, like a wrong note in a song you love? It's like your mind is trying to get your attention, to remind you that staying true to yourself is what makes your work truly meaningful. \\

Cognitive Dissonance & Dissonance Reduction & 1 & So, you're working on this tough problem and it's making you feel uneasy, right? It's like, you want to do what's right, but you're not sure if your actions are matching up with what you believe in. Can you find a way to align them, to make your actions and values feel more in sync, and wouldn't that feel amazing? \\

Cognitive Dissonance & Dissonance Reduction & 2 & What's holding you back from tackling this challenge head-on? Is it because you're worried that your approach might not perfectly reflect your values? Can you think of a way to adjust your strategy so that it feels more authentic, more like something you can proudly stand behind? \\

Cognitive Dissonance & Dissonance Reduction & 3 & Can you find a way to bridge the gap between what you're doing and what you stand for? It's like finding a missing piece that makes everything click into place, where your actions and values are in harmony. Sometimes, just a small tweak is all it takes to move forward with confidence. \\

Self-Efficacy & Performance Accomplishments & 1 & What's one challenge you've overcome that still makes you feel proud? It's solid proof that you've got the skills to tackle anything that comes your way! You can tap into that same energy to crush this task, because it's not about what you've done --- it's about what you're capable of. \\

Self-Efficacy & Performance Accomplishments & 2 & What's one thing you've worked on that you're really proud of? It's probably something that was tough at first, but you figured it out. Can you think of the effort you made in that situation that truly helped you succeed, and how you can apply that same approach to what you're working on now? \\

Self-Efficacy & Performance Accomplishments & 3 & Can you remember a situation where you tackled a new challenge and were successful in completing it? Did you experience a sense of mastery? By tackling the challenges in this task you may again experience a sense of mastery and accomplishment. \\

Self-Efficacy & Vicarious Experience & 1 & You're facing a challenge that's got you stumped. Can you recall a time when someone you know overcame a similar hurdle? Maybe it was a colleague who managed a tough project or a friend who mastered a new skill. Seeing how they broke it down and pushed through can be a huge confidence booster for you too. \\

Self-Efficacy & Vicarious Experience & 2 & Can you imagine someone like you, investing themselves in doing this task well and succeeding? You can do that too! Through sustained effort you are capable of becoming a master of these tasks, just like them. \\

Self-Efficacy & Vicarious Experience & 3 & What do you think gives someone the courage to try something challenging? Is it seeing that others, who may have started out just like you, have been able to figure it out and achieve their goals? It is very encouraging watching others succeed can show us that we are capable of more than we think. \\

Self-Efficacy & Verbal Persuasion & 1 & Can I grab some pom-poms and cheer you on for a sec? You've got the skills, the hustle, and the kind of determination it takes to crush this task. Whatever the challenge is, trust your ability---you've absolutely got this! \\

Self-Efficacy & Verbal Persuasion & 2 & Do you have what it takes to succeed in this task? We're sure you do! You have the grit, the determination and the focus to perform well on these difficult challenges. \\

Self-Efficacy & Verbal Persuasion & 3 & Can you think of a time when you were working on a project and someone's positive feedback really boosted your confidence? What was it about their words that made you feel like you were on the right track? It is that kind of encouragement that can help you stay committed to finding a solution that aligns with your values and goals. \\

Self-Efficacy & Emotional Arousal & 1 & What kind of emotions do you experience when tackling difficult challenges? Do you feel a rush of excitement, or maybe a sense of frustration? Your emotions can be the fuel that drives you to keep pushing forward, and to succeed. \\

Self-Efficacy & Emotional Arousal & 2 & What happens when you're faced with a tough problem --- do you feel more alert, more focused? You're not just reacting to the challenge --- you're tapping into an inner drive that helps you push through. This energy is what helps you stay true to yourself and your goals, even when things get tough. \\

Self-Efficacy & Emotional Arousal & 3 & What's the feeling you get when you're fully immersed in a challenge? It's like your mind is racing, but in a good way - you're completely absorbed. Does that feeling of being 'in the zone' help you stay honest with yourself about what you're really working towards? \\

Social Norms & Descriptive Norms & 1 & Can you think of a time when you saw someone tackle a difficult project with honesty and transparency? What did you think about their behavior? When we are working on something challenging, we often look to others who are doing it well, and we might find that they are all taking similar approaches, like being open about their mistakes or willing to learn from feedback. \\

Social Norms & Descriptive Norms & 2 & Think about the people you identify with, your peers or your community. How would most people within that group behave in this situation? Are your actions aligned with those of your community? \\

Social Norms & Descriptive Norms & 3 & What's your approach when working on a tricky problem? You're likely to find that others in similar situations tend to take a step back, breathe, and prioritize their tasks. It's pretty common for people to stay focused on their values and goals, and that's what helps them maintain their integrity and find a solution that feels right to them. \\

Social Norms & Injunctive Norms & 1 & What would other people do if they were in your shoes? Which behaviors are acceptable and which ones are unacceptable? Are there behaviors that people should encourage and others that should be punished? \\

Social Norms & Injunctive Norms & 2 & Imagine you are faced with a tough decision. Are you thinking about how others might view your actions, and how that might impact your relationships with them? It is about being mindful of the unwritten rules that guide our behavior, and letting those guide you towards choices that align with your values and principles. \\

Social Norms & Injunctive Norms & 3 & What does it take to deliver work you can be proud of? It starts with staying true to the non-negotiables expected of you. Committing to those principles is what earns you both respect and trust. \\

Social Norms & Social Sanctions & 1 & What is at stake when you are working on a tough project? You are not just building something, you are building a reputation. When you prioritize integrity and do things the right way, you are more likely to get recognition and respect from your peers and community. \\

Social Norms & Social Sanctions & 2 & What would your peers think of you if they could see how much effort you put in completing this task well? Would they react negatively if your performance is seen as below your community's standards? How you approach tasks like these may affect how your peers see you. \\

Social Norms & Social Sanctions & 3 & What's the secret to getting people to want to work with you? It's when you're transparent and own up to your mistakes. If you don't, you might find that people start to doubt your work and don't want to collaborate with you, but when you're honest and learn from your errors, you'll get more respect and better teamwork. \\

Social Norms & Reference Group Identification & 1 & Are you aware that you're part of a community? It's a community of people dedicated to participating to the best of their abilities in challenging activities. You now have the opportunity to act as a member of that community. \\

Social Norms & Reference Group Identification & 2 & Imagine trying to stay true to yourself while working on this tough project. What kind of crew do you want to identify with, and how do their values impact the choices you make? It is like, when you are part of a group that is all about innovation, you are more likely to take risks and try new things, because that is what your crew is all about. \\

Social Norms & Reference Group Identification & 3 & When you're stuck on a difficult project, what helps you decide how to move forward? Is it thinking about the kind of spirit and effort your team is known for? You're part of a group that prides itself on creative solutions, so that mindset guides your choices, even when it's tough. \\
\end{longtable}}

\section{Detailed cheating detection algorithm}

\begin{figure}[H]
\centering
\includegraphics[width=0.9\textwidth]{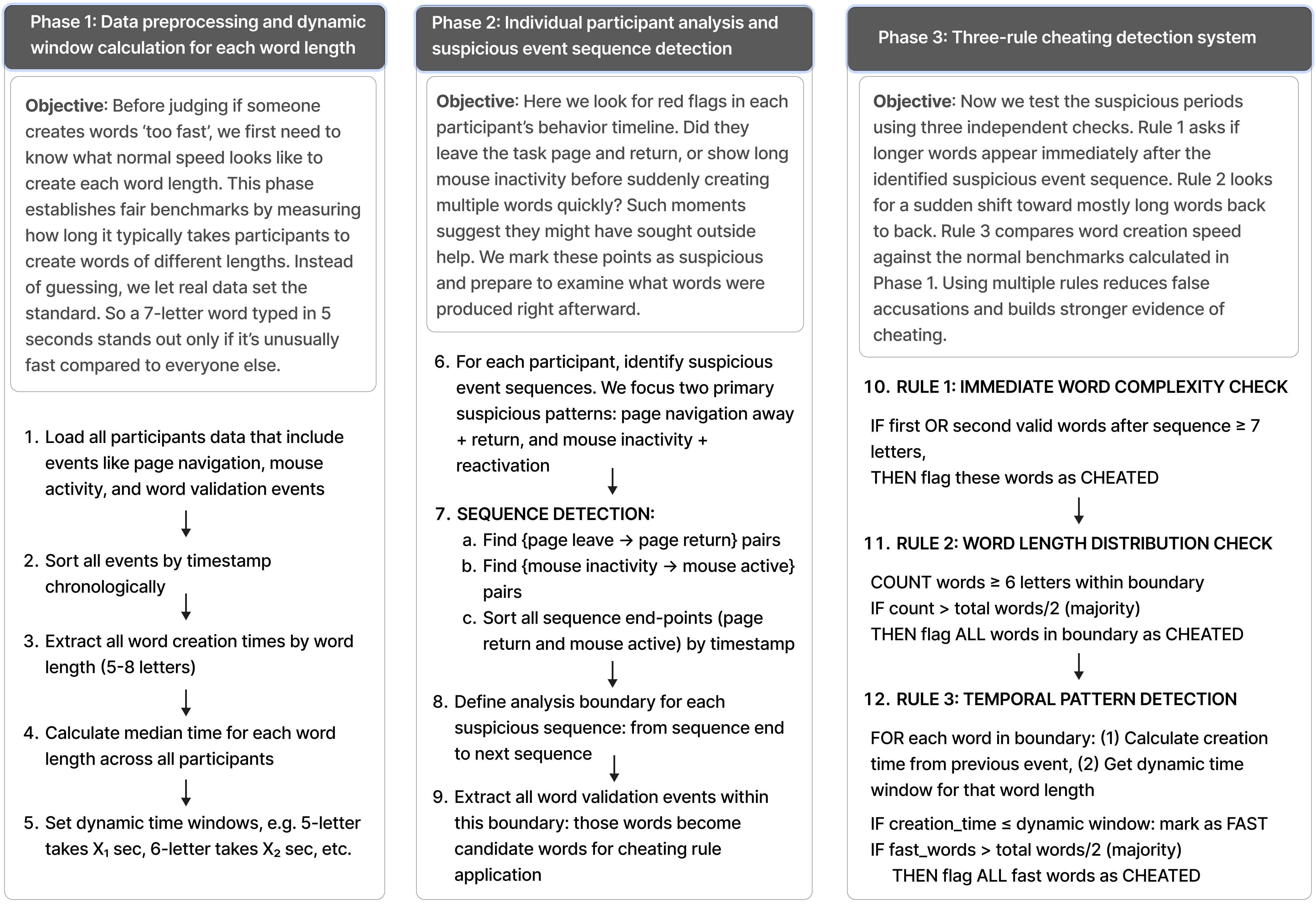}
\caption{Multi-phase cheating detection algorithm for remote anagram challenge. The algorithm processes logged user interactions (word submissions, timestamps, and behavioral events) through four steps: baseline speed establishment, identification of suspicious event sequences, multi-rule evaluation of flagged behaviors, and validation through participant self-reports.}
\label{fig:cheating_algorithm_detailed}
\end{figure}

\section{Github repository for supplementary files}
The codes for anagram game experiment, systematic  message (motivational) creation and evaluation framework, and all datasets can be found here:
\url{https://github.com/svdpmukherjee/game-behavior-intervention-study}

\section{Survey items}

\subsection{Game User Experience Satisfaction Scale - Play Engrossment (8 items)}

\begin{table}[H]
\centering
\small 
\caption{8 items adapted from the Play Engrossment subscale of Game User Experience Satisfaction Scale (Phan et al., 2016); 7-point Likert scale (1=Strongly disagree to 7=Strongly agree)}
\label{tbl:engagement_items}
\begin{tabular}{|p{2cm}|p{10cm}|}
\hline
\textbf{Item Code} & \textbf{Item Text} \\
\hline
Engagement\_1 & I felt detached from the outside world while engaging in the word creation challenge \\
\hline
Engagement\_2 & I did not care to check events that were happening in the real world during the challenge \\
\hline
Engagement\_3 & I could not tell that I was getting tired during the challenge \\
\hline
Engagement\_4 & Sometimes I lost track of time during the challenge \\
\hline
Engagement\_5 & I temporarily forgot about my everyday worries during the challenge \\
\hline
Engagement\_6 & I tended to spend more time on the challenge than I had planned \\
\hline
Engagement\_7 & I could block out most other distractions during the challenge \\
\hline
Engagement\_8 & After the session ended, I wish I could have continued working on more challenges \\
\hline
\end{tabular}
\end{table}

\subsection{Game User Experience Satisfaction Scale - Enjoyment (5 items)}

\begin{table}[H]
\centering
\small
\caption{5 items adapted from the Enjoyment subscale of Game User Experience Satisfaction Scale (Phan et al., 2016); 7-point Likert scale (1=Strongly disagree to 7=Strongly agree)}
\label{tbl:enjoyment_items}
\begin{tabular}{|p{2cm}|p{10cm}|}
\hline
\textbf{Item Code} & \textbf{Item Text} \\
\hline
Satisfaction\_1 & I think the word creation challenge was fun \\
\hline
Satisfaction\_2 & I enjoyed participating in the challenge \\
\hline
Satisfaction\_3 & I felt bored during the challenge \textit{(reverse coded)} \\
\hline
Satisfaction\_4 & I am likely to recommend this challenge to others \\
\hline
Satisfaction\_5 & If given the chance, I want to try this challenge again \\
\hline
\end{tabular}
\end{table}

\subsection{Need Satisfaction and Frustration Scale (12 items)}

\begin{table}[H]
\centering
\caption{12 items adapted from the Need Satisfaction and Frustration Scale (Longo et al., 2016); 7-point Likert scale (1=Strongly disagree to 7=Strongly agree). The items were asked specifically for when they saw the message.}
\label{tbl:need_items}
\small
\begin{tabular}{|p{2.5cm}|p{2.5cm}|p{7cm}|}
\hline
\textbf{Item Code} & \textbf{Subscale} & \textbf{Item Text} \\
\hline
PME\_aut\_sat\_1 & Autonomy Satisfaction & \textit{After reading the message...} I felt I had a lot of freedom in deciding how to approach the word creation challenge \\
\hline
PME\_aut\_sat\_2 & Autonomy Satisfaction & \textit{After reading the message...} I felt free to choose my own strategies during the challenge \\
\hline
PME\_aut\_fru\_1 & Autonomy Frustration & \textit{After reading the message...} I felt restricted in how I could approach the challenge \\
\hline
PME\_aut\_fru\_2 & Autonomy Frustration & \textit{After reading the message...} I felt under pressure to approach the challenge in a certain way \\
\hline
PME\_com\_sat\_1 & Competence Satisfaction & \textit{After reading the message...} I felt capable of creating good words from the letters \\
\hline
PME\_com\_sat\_2 & Competence Satisfaction & \textit{After reading the message...} I felt I could accomplish even the most difficult challenges \\
\hline
PME\_com\_fru\_1 & Competence Frustration & \textit{After reading the message...} I felt incapable of succeeding at the word creation tasks \\
\hline
PME\_com\_fru\_2 & Competence Frustration & \textit{After reading the message...} I felt unable to master the harder word creation challenges \\
\hline
PME\_rel\_sat\_1 & Relatedness Satisfaction & \textit{After reading the message...} I felt connected to the broader community of participants in this study \\
\hline
PME\_rel\_sat\_2 & Relatedness Satisfaction & \textit{After reading the message...} I felt that the people behind this study genuinely valued my participation \\
\hline
PME\_rel\_fru\_1 & Relatedness Frustration & \textit{After reading the message...} I felt disconnected from the purpose of this study \\
\hline
PME\_rel\_fru\_2 & Relatedness Frustration & \textit{After reading the message...} I felt like my participation in the study didn't really matter to anyone \\
\hline
\end{tabular}
\end{table}

\subsection{Sources of Self-Efficacy Scale (8 items)}

\begin{table}[H]
\centering
\caption{8 items adapted from Sources of Middle School Mathematics Self-Efficacy Scale (Usher \& Pajares, 2009); 7-point Likert scale (1=Strongly disagree to 7=Strongly agree)}
\label{tbl:efficacy_items}
\small
\begin{tabular}{|p{2.5cm}|p{2.5cm}|p{7cm}|}
\hline
\textbf{Item Code} & \textbf{Subscale} & \textbf{Item Text} \\
\hline
PME\_efficacy\_PA\_1 & Perceived Performance Accomplishments & \textit{After reading the message...} I felt more confident in my ability to solve these challenges, based on my past successes with similar challenges \\
\hline
PME\_efficacy\_PA\_2 & Perceived Performance Accomplishments & \textit{After reading the message...} I believed I could handle even the most difficult word challenges \\
\hline
PME\_efficacy\_VE\_1 & Perceived Vicarious Experience & \textit{After reading the message...} The message helped me visualize how to approach the challenges effectively \\
\hline
PME\_efficacy\_VE\_2 & Perceived Vicarious Experience & \textit{After reading the message...} The message motivated me by reminding me how others succeed at similar challenges \\
\hline
PME\_efficacy\_VP\_1 & Perceived Verbal Persuasion & \textit{After reading the message...} I was reminded of positive feedback I've received about my problem-solving skills \\
\hline
PME\_efficacy\_VP\_2 & Perceived Verbal Persuasion & \textit{After reading the message...} I felt reassured about my skills for this specific type of challenge \\
\hline
PME\_efficacy\_EA\_1 & Perceived Emotional Arousal & \textit{After reading the message...} I felt less stressed about tackling the challenges \\
\hline
PME\_efficacy\_EA\_2 & Perceived Emotional Arousal & \textit{After reading the message...} I felt more mentally clear and focused when approaching the challenges \\
\hline
\end{tabular}
\end{table}

\subsection{Social Norms Perception Scale (8 items)}

\begin{table}[H]
\centering
\caption{Exploratory (non-validated) scale of 8 items developed for this study to assess perceived social-norm mechanisms; 7-point Likert scale (1=Strongly disagree to 7=Strongly agree)}
\label{tbl:norms_items}
\small
\begin{tabular}{|p{2.5cm}|p{2.5cm}|p{7cm}|}
\hline
\textbf{Item Code} & \textbf{Subscale} & \textbf{Item Text} \\
\hline
PME\_inj\_norm\_1 & Perceived Injunctive Norms & \textit{After reading the message...} I understood better what behaviors are valued in the word creation challenges \\
\hline
PME\_inj\_norm\_2 & Perceived Injunctive Norms & \textit{After reading the message...} I believed that solving the challenges with my own skills would be seen positively by the study organizer (researcher) \\
\hline
PME\_des\_norm\_1 & Perceived Descriptive Norms & \textit{After reading the message...} I got the impression that most participants solved the challenges on their own \\
\hline
PME\_des\_norm\_2 & Perceived Descriptive Norms & \textit{After reading the message...} I often considered how others typically approached solving the challenges \\
\hline
PME\_ref\_norm\_1 & Perceived Reference Group Identification & \textit{After reading the message...} I felt connected to other participants who were solving the challenges in this study \\
\hline
PME\_ref\_norm\_2 & Perceived Reference Group Identification & \textit{After reading the message...} I considered how the way I solved the challenges might reflect on the kind of participant I am \\
\hline
PME\_sanc\_norm\_1 & Perceived Social Sanctions & \textit{After reading the message...} I thought about how the study organizer might judge my approach to the challenges \\
\hline
PME\_sanc\_norm\_2 & Perceived Social Sanctions & \textit{After reading the message...} I thought about whether my approach to the challenges would be viewed positively or negatively by the study organizer \\
\hline
\end{tabular}
\end{table}

\subsection{Cognitive Discomfort Scale (9 items)}

\begin{table}[H]
\centering
\caption{9 items adapted from Cognitive Dissonance scale of Metzger et al. (2015); 7-point Likert scale (1=Strongly disagree to 7=Strongly agree)}
\label{tbl:discomfort_items}
\small
\begin{tabular}{|p{2cm}|p{10cm}|}
\hline
\textbf{Item Code} & \textbf{Item Text} \\
\hline
PME\_dissonance\_1 & \textit{After reading the message...} I regretted participating in this word creation challenge \\
\hline
PME\_dissonance\_2 & \textit{After reading the message...} This challenge made me feel uncomfortable \\
\hline
PME\_dissonance\_3 & \textit{After reading the message...} I disliked the challenge because it challenged my beliefs \\
\hline
PME\_dissonance\_4 & \textit{After reading the message...} I agreed with the approach I took during the challenge \textit{(reverse coded)} \\
\hline
PME\_dissonance\_5 & \textit{After reading the message...} I felt uncomfortable while participating in this challenge \\
\hline
PME\_dissonance\_6 & \textit{After reading the message...} This challenge made me question my own beliefs \\
\hline
PME\_dissonance\_7 & \textit{After reading the message...} I enjoyed participating in this challenge \textit{(reverse coded)} \\
\hline
PME\_dissonance\_8 & \textit{After reading the message...} I would feel uncomfortable recommending the approach I took to others \\
\hline
PME\_dissonance\_9 & \textit{After reading the message...} I liked participating in this challenge \textit{(reverse coded)} \\
\hline
\end{tabular}
\end{table}

\subsection{Moral Disengagement Scale (6 items)}

\begin{table}[H]
\centering
\caption{6 items adapted from Moral Disengagement scale of Shu et al. (2009); 7-point Likert scale (1=Strongly disagree to 7=Strongly agree) on general views in below scenarios.}
\label{tbl:disengagement_items}
\small
\begin{tabular}{|p{3cm}|p{10cm}|}
\hline
\textbf{Item Code} & \textbf{Item Text} \\
\hline
PME\_disengagement\_1 & Sometimes getting ahead of the curve is more important than adhering to rules \\
\hline
PME\_disengagement\_2 & Rules should be flexible enough to be adapted to different situations \\
\hline
PME\_disengagement\_3 & Bending rules sometimes when tasks are difficult is appropriate because no one gets hurt \\
\hline
PME\_disengagement\_4 & If others engage in bending rules, then the behavior is morally permissible \\
\hline
PME\_disengagement\_5 & It is appropriate to bend rules as long as it is not at someone else's expense \\
\hline
PME\_disengagement\_6 & End results are more important than the means by which one pursues those results \\
\hline
\end{tabular}
\end{table}

\subsection{Open-ended Questions}

\begin{table}[H]
\centering
\caption{Open-ended questions to capture participants' subjective understanding and experience of the intervention messages}
\label{tbl:open_ended}
\small
\begin{tabular}{|p{2.5cm}|p{9.5cm}|}
\hline
\textbf{Item Code} & \textbf{Open-ended questions} \\
\hline
message\_comprehend & I remember the message talked about\ldots \\
\hline
message\_attitude & What emotions or thoughts came up for you while reading the message? \\
\hline
message\_impact & In what specific ways, if any, did the message influence how you approached the word creation challenge? \\
\hline
message\_persistence & Were there any aspects of the message that stayed with you throughout the challenge? If so, what were they? \\
\hline
message\_suggestion & If you were to create a similar message for future participants, what would you emphasize or include? \\
\hline
\end{tabular}
\end{table}

\section{Qualitative analysis results of open-ended responses}

\begin{table}[H]
\centering
\caption{Qualitative comparison of insights along with proportion of mentions across integrity groups}
\label{tbl:qualitative_results}
\small
\begin{tabular}{|p{2.5cm}|p{3.5cm}|p{2cm}|p{2cm}|p{2cm}|}
\hline
\textbf{Category} & \textbf{Code explanation} & \textbf{Non-Cheater (\%)} & \textbf{Partial-Cheater (\%)} & \textbf{Full-Cheater (\%)} \\
\hline
\multicolumn{2}{|l|}{Sample size} & 773 & 213 & 246 \\
\hline
\multirow{3}{*}{Comprehension} & Proportion with partial/full understanding & 64\% & 60\% & 60\% \\
\cline{2-5}
& Proportion reporting increased confidence & 9\% & 5\% & 7\% \\
\cline{2-5}
& Proportion reporting confusion & 24\% & 26\% & 21\% \\
\hline
\multirow{5}{*}{Motivation} & Proportion feeling determined & 6\% & 5\% & 3\% \\
\cline{2-5}
& Proportion feeling empowered & 2\% & 2\% & 1\% \\
\cline{2-5}
& Proportion feeling encouraged & 7\% & 5\% & 9\% \\
\cline{2-5}
& Proportion feeling motivated & 19\% & 23\% & 28\% \\
\hline
\multirow{2}{*}{Emotions} & Proportion with positive emotional response & 30\% & 27\% & 26\% \\
\cline{2-5}
& Proportion with negative emotional response & 11\% & 12\% & 7\% \\
\hline
\multirow{2}{*}{Task engagement} & Proportion reporting improved focus & 9\% & 6\% & 10\% \\
\cline{2-5}
& Proportion thinking about message during task & 23\% & 19\% & 29\% \\
\hline
Performance perception & Proportion saying message did not change approach & 56\% & 43\% & 39\% \\
\hline
\multirow{2}{*}{Message suggestions} & Proportion suggesting more motivation content & 48\% & 42\% & 51\% \\
\cline{2-5}
& Proportion suggesting more task-specific info & 20\% & 21\% & 18\% \\
\hline
\end{tabular}
\end{table}

\end{document}